\documentclass[usenatbib]{mnras}
\usepackage{newtxtext,newtxmath}
\usepackage[T1]{fontenc}
\usepackage{ae,aecompl}
\usepackage{graphicx}
\usepackage{amsmath}
\usepackage{scalerel}

\usepackage[dvipsnames]{xcolor}

\newcommand{\vecv}{\mbox{\boldmath $v$} {}}
\newcommand{\vece}{\mbox{\boldmath $e$} {}}
\title[Gaps by planets in eccentric orbit]
{Estimating the depth of gaps opened by planets in eccentric orbit
}

\author[S\'anchez-Salcedo et al.]{
F. J. S\'anchez-Salcedo,$^{1}$\thanks{E-mail:jsanchez@astro.unam.mx} R. O. Chametla$^{2}$ and O. Chrenko$^{2}$
\\
$^{1}$Instituto de Astronom\'{\i}a, Universidad Nacional Aut\'{o}noma de M\'{e}xico, AP 70-264, 
Mexico City 04510, Mexico  \\   
$^{2}$Charles University, Faculty of Mathematics and Physics, Astronomical Institute, V Hole\v{s}ovi\v{c}k\'ach 747/2, 180 00 Prague 8, Czech Republic
}

\date{Accepted XXX. Received YYY; in original form ZZZ}

\pubyear{2015}

\begin{document}
\label{firstpage}
\pagerange{\pageref{firstpage}--\pageref{lastpage}}
\maketitle

\begin{abstract}

Planets can carve gaps in the surface density of protoplanetary discs.
The formation of these gaps can reduce the corotation torques acting on the planets. In addition,
gaps can halt the accretion of solids onto the planets 
as dust and pebbles can be trapped at the edge of the gap. 
This accumulation of dust could explain the origin of the ring-like dust 
structures observed using high-resolution interferometry. In this work 
we provide an empirical scaling relation for the depth of the gap cleared by
a planet on an eccentric orbit as a function of the planet-to-star mass ratio $q$,
the disc aspect ratio $h$, Shakura-Sunyaev viscosity parameter $\alpha$, and planetary
eccentricity $e$. We construct the scaling relation using a heuristic approach: we calibrate
a toy model based on the impulse approximation with 2D hydrodynamical simulations.
The scaling reproduces the gap depth for moderate eccentricities ($e\leq 4h$)
and when the surface density contrast outside and inside the gap is
$\leq 10^{2}$. Our framework can be used as the basis of more sophisticated models aiming to predict
the radial gap profile for eccentric planets.

\end{abstract}

\section{Introduction}
The gravitational interaction between a planet and its natal protoplanetary disc leads
to a transfer of angular momentum between them. For isothermal disc models,
the tidal torques on the planets cause them to migrate inward. The waves excited by the 
planet deposit angular momentum in the disc and thus the density structure of the disc may be modified.
In particular, planets that exceed a certain critical mass open a depleted gap in the disc
\citep[e.g.][]{lin93}. 
This change in the density structure of the disc is relevant to understand
the migration rate of gap-opening planets because the torque on the planets depends
on the surface density in the gap \citep{lin86,war97,kan18}. 
In addition, the gas accretion
rate onto the planet strongly depends on the gas surface density in the gap 
\citep[e.g.,][]{bry99,tan16}.
Moreover, the presence of a gap produces a radial pressure bump which can halt the accretion 
of pebbles onto the planet \citep{hag03,pin12}.
Gaps/rings in the dust in protoplanetary discs can be directly observed by the
Atacama Large Millimeter Array (ALMA) in mm or sub-mm wavelengths 
\citep[e.g.,][]{guz18,hua18,lon18,lod19}.

A great effort has been done to characterize the profile (width and depth) of 
planet-induced gaps \citep{tak96,bry99,bat03,var04,cri06,duf13,fun14,kan15,kan17,duf15a,cha17,don17,gin18,dem20,duf20,tan22}.
Using numerical simulations, \citet{duf13} find a simple empirical relation between
the depth of the (gas) gap carved by a planet in circular orbit and the planet-to-star mass ratio 
$q$, the disc aspect ratio $h$, and the disc viscosity parameter $\alpha$.
\citet{fun14} estimate the gap depth using a ``zero-dimensional''
model. They assume that the one-sided Lindblad torque is 
$\approx q^{2} h^{-3}\Sigma_{\rm gap}\omega^{2} a^{4}$, where $\Sigma_{\rm gap}$ is the surface
density at the bottom of the gap, $\omega$ is the angular frequency 
of the planet, and $a$ its orbital radius. By balancing the
Lindblad torque with the viscous torque, they obtain
the simple scaling $\Sigma_{\rm gap}\propto q^{-2} \alpha h^{5}$.
\citet{kan15} and \citet{duf15a}  
study the gap profile adopting a
one-dimensional model. They also assume that the excitation torque is determined by
the Lindblad torque but take into account that the deposition of angular momentum is 
not local (instantaneous). 
The model of \citet{duf15a} was successful in explaining the depth and width 
of partial gaps ($\Sigma_{\rm gap}/\Sigma_{0}\gtrsim 0.6$), where the waves launched 
by the planet are weakly nonlinear. 
\cite{gin18} derive the deposition locations of density waves in the
presence of a deep gap. 
For low-mass planets and low viscosities, they demonstrate 
that besides the classical maximum at a radial distance $ha$, the torque density exhibits
a second peak,  forming a ``two-step'' density profile.
For high values of $q$ (deep gap), \citet{kan17} and \citet{duf20} 
construct empirical formulae for the surface density distributions of the gap, which
reproduce the results of 2D simulations.

Less well developed is the description of the gap opened by a planet in eccentric orbit.
Most of the studies of the interaction between an eccentric planet and the disc have
focused on the planetary orbital evolution, either using an analytical approach
\citep[e.g.,][]{pap00,gol03,tan04,moo08,mut11,tey16} or hydrodynamical simulations 
\citep[e.g.,][]{cre06,cre07,bit10,bit13,fen14,duf15b,rag18,san19,leg21}.

The shape of the gap formed by planets on eccentric orbits ($e\leq 0.2$) was 
investigated by \citet{hos07}.
They find that if the eccentricity is below $(q/3)^{1/3}$, the gap
is almost identical to the circular case. As the eccentricty increases above this
value, the gap, as observed in the azimuthally averaged surface density profile, 
becomes wider and shallower.
This was attributed to the fact that the corotation torque decreases as
$e$ increases. 
\citet{hos07} also suggest a criterion for gap formation by eccentric planets.

\citet{duf15b} conduct 2D simulations of a planet on an eccentric
orbit in order to explore under which conditions the eccentricity can be excited.
As a by-product of their simulations, they report how the gap depth depends on 
the eccentricity in the range $0\leq e \leq 0.12$; the gap becomes shallower as the 
eccentricity increases.

The gap may also affect the rate of dust accretion onto the planet
because the outer edge of the gap acts as a pressure bump which can block 
the radial drift of solids, mainly
pebbles. When the pressure bump formed by a planet becomes strong enough to render the
gas orbital velocity super-Keplerian, it starts to act as 
a barrier against the radial drift of pebbles and the planet stops accreting
them \citep[][]{M_N2012,L_M2014,Ataiee+2018,Bitsch+2018}.
If the latter occurs, the planet is said to have reached its Pebble Isolation Mass (PIM).
Recently, \citet{Chametla2022} have studied the PIM for a planet on a fixed eccentric orbit for planetary eccentricities up to $e=0.2$ and $\alpha-$turbulent viscosities between $10^{-4}$ and $10^{-2}$. They use gas-dust 2D simulations, including dust turbulent diffusion,
and find that eccentric planets reach a well-defined PIM, which can be smaller than it is for planets on circular orbits as long as $e\leq h$.

The problem of the disc-planet interaction for eccentric orbits has renewed
interest by the recent observations that circumstellar discs display spiral arms, 
dust rings and gaps. \citet{li19} describe the gas and
dust distributions, during and after circularization, induced by a planet of $10 M_{\oplus}$ 
with an initial eccentricity of $e=0.8$. \citet{cal20} find that an unseen planet of
$9.5M_{J}$ with an eccentricity of $e\simeq 0.4$ can reproduce several features observed in
MWC 758 protoplanetary disc.
Based on simulations of discs with gas and dust,  \citet{che21} compare the dust gap opened
by planets with $q=(1-3)\times 10^{-4}$ in circular and eccentric ($e=0.1$) orbits. 
They show that the dust gap is wider for
eccentric planets provided that their radial epicyclic excursion is larger than their 
Hill radius. Therefore, there is a mass-eccentricity degeneracy in the sense that
the width of the dust gap does not determine the planet mass uniquely.
On the other hand, \citet{zhu22} develop a method to predict the location of
spiral arms excited by eccentric ($e\leq 0.5$) low-mass planets. They suggest that a planet in eccentric orbit could explain the formation of multiple spiral arms observed in some systems (e.g., in HD 34700A), and their large pitch angles. If the mass of the planet is known, it could be viable to constrain the eccentricity of a planet from measurements of the pattern speed of spiral arms.

The main objective of this paper is to derive an analytical formula for the scaling 
of the gap depth with $q$, $h$, $\alpha$ and $e$. 
The paper is organized as follows. 
Our semi-analytical framework to 
model the gaps carved by eccentric planets is presented in Section \ref{sec:model_sec2}. 
The scaling of the gap depth is determined empirically in Section \ref{sec:num_models},
where we calibrate our model by comparing with the results of numerical simulations.
In Section \ref{sec:ecc_damping} we give a brief review on the level of eccentricity
of planets embedded in protoplanetary discs.
Finally, our main conclusions are given in Section \ref{sec:conclusions}.

\section{Gaps produced by eccentric planets: theoretical estimates}
\label{sec:model_sec2}
\subsection{Gaps produced by planets: Overview}
\label{sec:overview}

Consider a planet with semimajor axis $a$, eccentricity $e$ and mass $M_{p}$,
on an orbit around a star of mass $M_{\star}$.  The planet is 
embedded in a disc with unperturbed surface density $\Sigma_{0}(R)$ and pressure scale height $H(R)$.
In the presence of the planet, the surface density of the disc is modified.
One-dimensional models of the gap assume that the 
surface density is approximately axisymmetric; $\Sigma =\Sigma(t,R)$. 

The surface density of an axisymmetric viscous disc evolves according to
\begin{equation}
\frac{\partial \Sigma}{\partial t} = \frac{1}{R} \frac{\partial}{\partial R}
\left( \frac{3}{\Omega R} \frac{\partial}{\partial R} (\nu \Omega R^{2} \Sigma)-
\frac{\Lambda_{\rm dep}}{\pi \Omega R}\right),
\label{eq:Sigma_time_1}
\end{equation}
where $\nu$ is the disc kinematic viscosity,  $\Omega(R)$ the angular frequency of the disc
and $\Lambda_{\rm dep}(R)$ is the deposition torque density (per unit radius)
\citep[e.g.,][]{tak96,iva15}.  
In order to compute the deposition torque,
one first needs the  excitation torque exerted on the disc by the planet.

The excitation torque density by a planet in circular orbit was computed by \citet{gol80}.
They assume that the planet is not massive enough to open a gap [so that $\Sigma(R)\simeq \Sigma_{0}(R)$]
and that $\Sigma_{0}$ is uniform on scales $\sim |R-a|$. 
  \citet{gol80} find that at $|R-a|>H_{a}$, where $H_{a}$ is the disc scale height at $R=a$,
the excitation torque density (per unit radius) exerted by a planet with $e=0$ is 
\begin{equation}
\Lambda_{\rm exc} (R) \simeq 
 0.8\pi q^{2}\Sigma_{0} a^{3}\omega^{2} \left(\frac{a}{R-a}\right)^{4} {\rm sign} (R-a),
\label{eq:density_GT80}
\end{equation}
where $q\equiv M_{p}/M_{\star}$ and $\omega^{2}=GM_{\star}/a^{3}$.
In the region $|R-a|<H_{a}$, the torque density dramatically decreases with decreasing
 $|R-a|$. 

The one-sided torque $T_{1s}$ can be evaluated by integrating the torque density
from the cut-off distance $H_{a}$ to infinity. Doing so we find
\begin{equation}
T_{1s} = 0.8 q^{2} \Sigma_{0} a^{4}\omega^{2} h^{-3}.
\end{equation}

For gap-opening planets, 
the one-sided excitation torque $T_{1s}$ is given by the formula inferred by \citet{gol80}  
but $\Sigma_{0}$ is replaced with the surface density $\Sigma_{\rm gap}$ at the
bottom of the gap
\begin{equation}
T_{1s}=f_{0} q^{2} \Sigma_{\rm gap} a^{4} \omega^{2} h^{-3},
\label{eq:Duffell_exact}
\end{equation}
where $f_{0}$ is a constant. 
\citet{duf15a} derives that $f_{0}=0.45$ by comparing analytic predictions with simulations.

The same scaling for the excitation torque can be derived also in the 
impulse (non-resonant) approximation:
\begin{equation}
\Lambda_{\rm exc}(R) = \frac{8}{9}q^{2}\Sigma_{\rm gap} 
a^{3}\omega^{2} \left(\frac{a}{R-a}\right)^{4} {\rm
sign}(R-a),
\label{eq:circular_density_tq}
\end{equation}
\citep[e.g.,][]{lin79,lin93,arm10,lub10}.
In order to match the torque found by \citet{duf15a},
it is needed to introduce a cut-off length at $R_{\rm cut}=0.87H_{a}$.

For planets in circular orbit, the deposition torque density $\Lambda_{\rm dep}$ is
computed using a model that describes how the waves excited by the planet steepen
into a shock. When shocks form, dissipation becomes effective and the angular momentum
can be deposited in the disc \citep{goo01,raf02,duf15a}.
The analysis is based on
the assumption that the excitation region is separated from the damping region
(excitation occurs at resonances near the planet, whereas damping occurs at larger
distances). For planets on a circular orbit, there is a clear separation between the outer
disc ($R>a$) and the inner disc ($R<a$) in the sense that all the positive (negative) torque excited
at the outer (inner) disc is deposited in the outer (inner) disc.

For eccentric planets, the situation is more
complex; the waves are excited at different radial distances and exhibit different
pitch angles depending on the planetary orbital phase. 
The torque excited at the outer disc can be deposited in the inner disc and vice versa. 
The flow is so complex that,
to our knowledge, the excitation torque density 
by an eccentric planet has only been found through numerical simulations \citep{bit10}.

\subsection{The excitation torque density in the local non-resonant approximation}
\label{sec:tq_density}

We assume that
the orbital eccentricity remains small enough so that the planet-induced gap
does not substantially deviate from axial symmetry and thus it is reasonable to adopt one-dimensional models to describe its profile $\Sigma (R)$. In practice, $\Sigma(R)$ should be thought of as the azimuthally averaged surface density. 

Following \citet{san19}, we adopt a system of reference where the
pericentre of the planet is at $x=(1-e)a$, $y=0$ and $z=0$. Therefore, the distance of
the planet to the central object $R_{p}$, and the planet's velocity $\vecv_{p}$ are 
\begin{equation}
R_{p} = \frac{\eta^{2} a}{1+e\cos\theta},
\label{eq:Rp}
\end{equation}
and
\begin{equation}
\vecv_{p} = \eta^{-1} \omega a (e\sin \theta \,\vece_{R} + (1+e\cos\theta)\,
\vece_{\theta}),
\end{equation}
where $\eta = \sqrt{1-e^{2}}$, $\theta$ is the true anomaly, $(\vece_{R},\vece_{\theta})$ are the unit
vectors, and we recall that $\omega$ is
the mean motion of the planet.
On the other hand, the unperturbed orbital velocity of the gas is
\begin{equation}
\vecv_{g} (R) = R \Omega \sqrt{ 1+\frac{1}{\Sigma R\Omega^{2}} \frac{dP}{dR}}
\vece_{\theta},
\label{eq:vg}
\end{equation}
where $\Omega (R) = \sqrt{GM_{\star}/R^{3}}$ and $P$ is the gas pressure.

We define the relative velocity $\vecv_{\rm rel}$ of the perturber relative to a gas streamline with radius $R$.
For small eccentricities $e\lesssim 0.2$, we approximate
\begin{equation}
\vecv_{\rm rel} \simeq a \left(\frac{v_{p,\theta}}{R_{p}} - \frac{v_{g}}{R}\right) \vece_{\theta},
\label{eq:vrelA}
\end{equation}
where $v_{p,\theta}$ is the azimuthal velocity of the planet
\citep[e.g.,][]{lin79,pap06,lub10}.
In the above equation, the radial component of the velocity of the planet has been 
ignored. This is
justified for small eccentricities. For instance, for $e=0.16$, the velocity of the planet
is underestimated by $1.3\%$ and the maximum pitch angle of the orbit is $9.2^{\circ}$.

\begin{figure*}
\includegraphics[width=188mm, height=224mm]{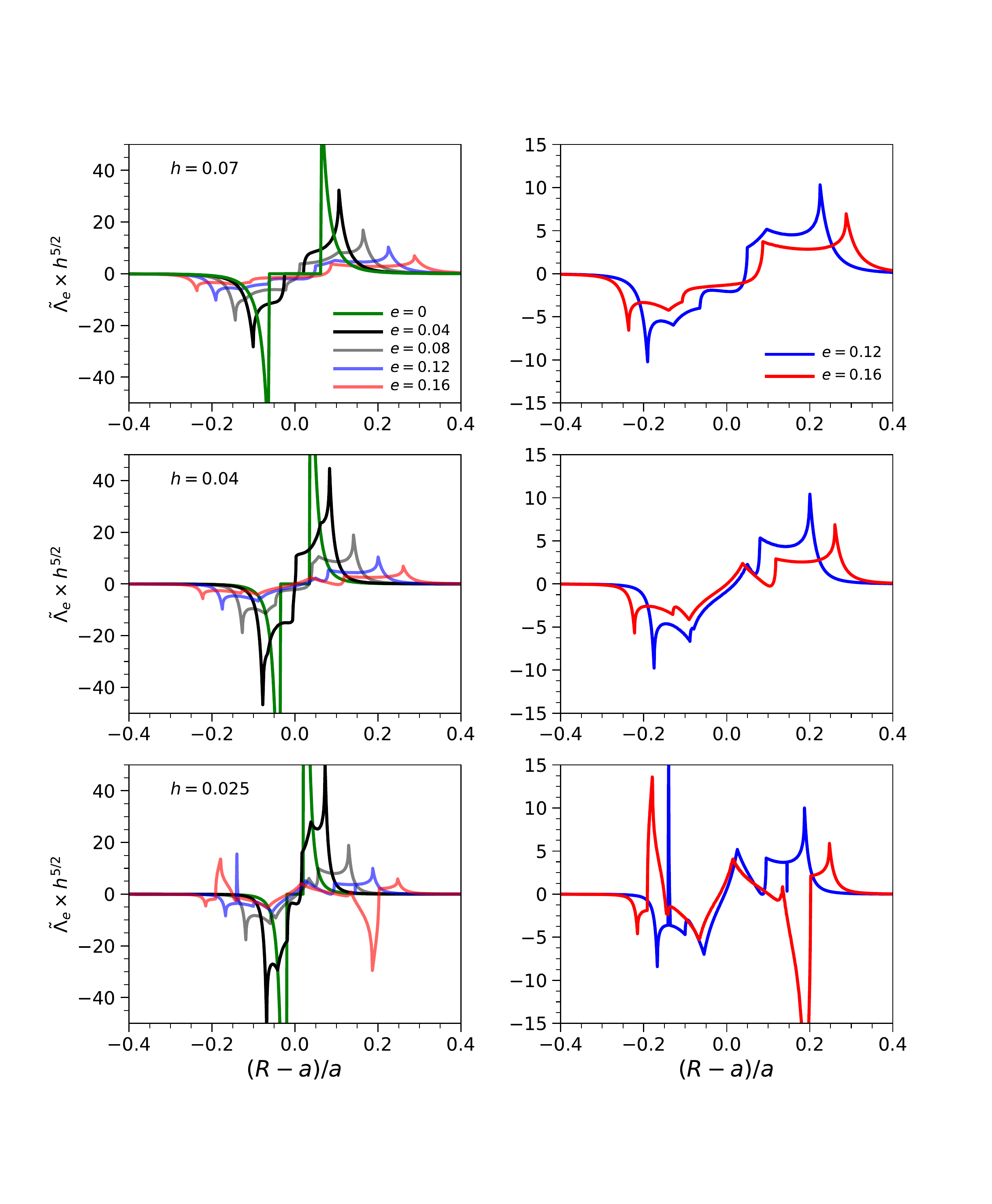}
\vskip -0.3cm
\caption{Torque density $\tilde{\Lambda}_{e}$ (multiplied by $h^{5/2}$)
for the aspect ratio $h=0.07$ (upper panels), $0.04$ (middle panels)
 and $0.025$ (lower panels). Individual curves correspond to calculations for different orbital eccentricities $e$ (see the legends 
 in bottom right corners of the upper panels). In the right panels, the vertical axis has been rescaled to improve legibility of the cases $e=0.12$ and $0.16$. We adopted $R_{\rm cut}=0.9H_{p}$. We scale the torque density
 by $h^{5/2}$ in order to have the same vertical axis range, which facilitates comparison between
 cases with different $h$.
 }
\label{fig:torque_density}
 \end{figure*}

Let us denote $\Delta\equiv R-R_{p}$ the impact parameter between a streamline and the perturber.
Expressing $v_{g}$ from Equation (\ref{eq:vg}) in terms of $\Delta$, 
we have
\begin{equation}
\frac{v_{g}}{R}= \sqrt{\frac{GM_{\star}}{(R_{p}+\Delta)^{3}}}=
\omega\left(\frac{a}{R_{p}+\Delta}\right)^{3/2},
\label{eq:vgR}
\end{equation}
where we have omitted the pressure gradient. This is justified for thin discs
in which the deviation from Keplerian rotation due to pressure support
is $\simeq 1 \%$.
Substituting Equation (\ref{eq:Rp}) into Equation (\ref{eq:vgR}), we can write
\begin{equation}
\frac{v_{g}}{R}\simeq \eta^{-3}\omega \xi^{3/2} \left( 1- \frac{3}{2\eta^{2}}
\frac{\Delta}{a}\xi\right),
\label{eq:vgR_over_R}
\end{equation}
where $\xi\equiv 1+e\cos\theta$. Combining Eqs. (\ref{eq:vrelA}) and
(\ref{eq:vgR_over_R}), and using $v_{p,\theta}/R_{p}=\eta^{-3}\omega \xi^{2}$, we obtain that
$\vecv_{\rm rel}=v_{\rm rel} \vece_{\theta}$ with
\begin{equation}
v_{\rm rel} \simeq \eta^{-3}\omega a \xi\left(\xi -\sqrt{\xi}
+\frac{3}{2\eta^{2}} \frac{\Delta}{a} \xi^{3/2} \right).
\label{eq:vrel_mod}
\end{equation}
Note that $v_{\rm rel}$ is not the modulus of the relative velocity;  it is negative
if the disc particle orbits faster relative to the planet. For instance, consider
a planet in circular orbit ($e=0$). In such a case,
$\xi=1$ and $v_{\rm rel}= 3\omega \Delta/2$. Hence
$v_{\rm rel}$ is negative at $\Delta <0$, i.e. in the disc region inwards from the planet.

\begin{figure}
\includegraphics[width=85mm, height=148mm]{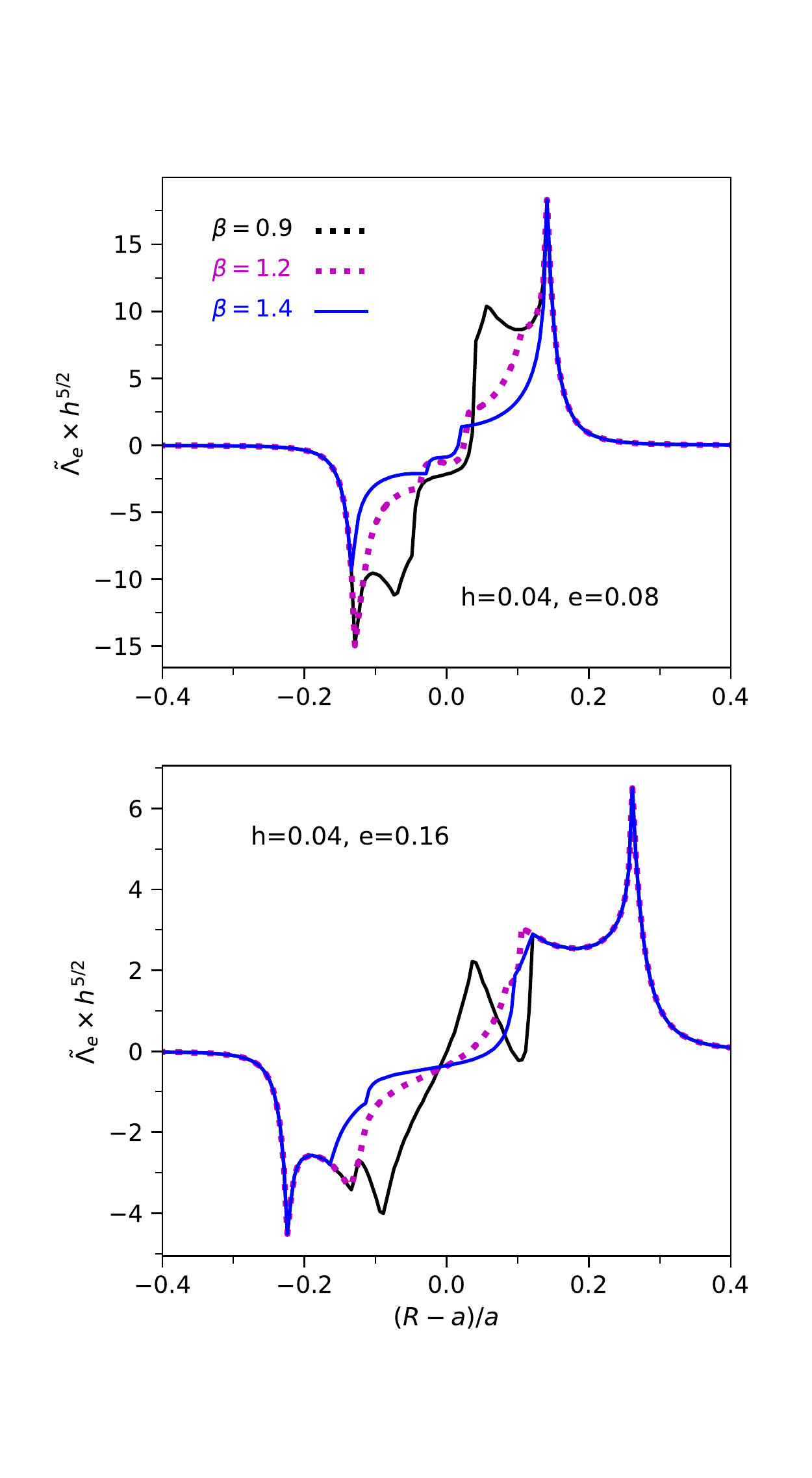}
\vskip -1.0cm
\caption{Torque density $\tilde{\Lambda}_{e}$ (multiplied by $h^{5/2}$)
for three different values of $\beta$ ($0.9, 1.2$ and $1.4$).}
\label{fig:tq_diff_Rsoft}
 \end{figure}

In a passage with the impact parameter $\Delta$, a disc particle of unit mass
gains (or loses) an amount of angular momentum $\delta h_{e}$ given by 
\begin{equation}
\delta h_{e} = \frac{2\varepsilon G^{2}M_{p}^{2}R_{p}}{v_{\rm rel}^{3} \Delta^{2}},
\label{eq:gas_scatter}
\end{equation}
where
\begin{equation}
\varepsilon= 
\left\{
\begin{array}{cl}
1 &\textup{if  } |v_{\rm rel}|>c_{s} \,\,
\textup{and } \,|\Delta|>R_{\rm cut} \\
  &  \\
0 & \textup{otherwise}
\end{array}
\right.
\label{eq:epsilon_def}
\end{equation}
\citep[e.g.][]{lin79,pap06}.
Here we have assumed that the change of angular momentum of a gas particle is
zero if the relative velocity is smaller than the local sound speed $c_{s}$,
The reason is that the free-streaming approximation implicit in 
Equation (\ref{eq:gas_scatter}) is not appropriate when the relative velocity is subsonic. 
We have also introduced
$R_{\rm cut}$ as the minimum impact parameter for which our
2D approximation, which ignores the disc thickness,
is no longer valid. This occurs when 
$\Delta$ is comparable to or lower than the vertical scale height. 
Therefore, we take $R_{\rm cut}\simeq \beta H_{p}$, with $H_{p}\equiv H(R_{p})$ being
the scale height of the disc
at the instantaneous position of the planet and
$\beta$ being a factor of order unity.

The transfer rate of the angular momentum to an annulus at a given $x\equiv R-a$,
i.e. the excitation torque density at $x$, is obtained by integrating over one planetary orbit:
\begin{equation}
\Lambda_{e}(x) = \Sigma(x) 
\int_{1-e}^{1+e}\mathcal{P} (\xi) \,\delta h_{e} \,|v_{\rm rel}|
\, d\xi,
\label{eq:torque_density0main}
\end{equation}
where $\mathcal{P}(\xi)\,d\xi$ is the fraction of the orbital period that the planet
spends between $\xi$ and $\xi +d\xi$. In an eccentric ($e\neq 0$) Keplerian orbit, it is
\begin{equation}
\mathcal{P}(\xi)= \frac{1}{\mathcal{I}_{e}\xi^{2}\sqrt{e^{2}-(\xi-1)^{2}}},
\label{eq:time_fraction}
\end{equation}
where
\begin{equation}
\mathcal{I}_{e}= \int_{0}^{\pi} \frac{d\theta}{(1+e\cos\theta)^{2}}\simeq \pi\left(1+\frac{3e^{2}}{2}\right).
\end{equation}
More details about the computation of the torque density
are given in Appendix \ref{sec:explicit_form}.

\begin{figure*}
\hskip 1.5cm
\includegraphics[width=190mm, height=70mm]{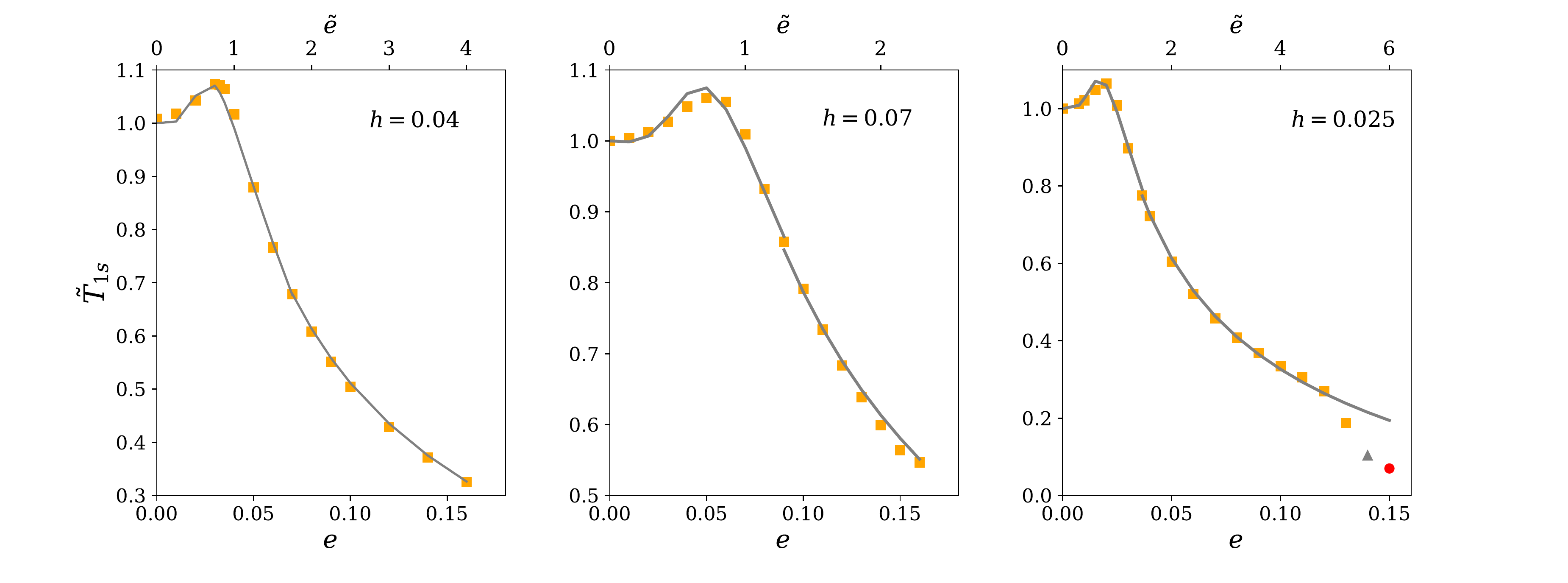}
\caption{One-sided torque normalized to the one-sided torque in the circular case,
as a function of the eccentricity. Individual panels correspond to different values of $h$.
Results of our calculations, using $\beta=0.9$ as described in Section \ref{sec:tq_density},
are shown by symbols whereas the gray curves are empirical fits (see Equation \ref{eq:fitT1s}).
Squares indicate that $|\lambda|<0.05$. Only two points
in the right panel have a larger value of $|\lambda|$: the black triangle 
($|\lambda|=0.38$) and the red dot ($|\lambda|=1$).
}
\label{fig:tq_onesided}
\end{figure*}

Figure \ref{fig:torque_density} shows the torque density (per unit of radial distance
and per unit of surface density)
\begin{equation}
\tilde{\Lambda}_{e}(x)\equiv \frac{\Lambda_{e}(x)}{q^{2} \Sigma(x) a^{3}\omega^{2}},
\end{equation}
 in discs having
a constant aspect ratio $h=H/R=\mathrm{const}$\footnote{Indeed, we will take $h=\mathrm{const}$ in
all models presented in this paper.}, for different combinations
of $h$ and $e$. 
Note that $h$ determines the disc sound speed, which is required to compute
$\varepsilon$ in Equation (\ref{eq:epsilon_def}).
We assumed that $R_{\rm cut}=0.9H_{p}$, i.e. $\beta=0.9$.
This value was not chosen arbitrarily; it is the value found in Section \ref{sec:overview} 
that leads to a correct evaluation of the one-sided torque for $e=0$
and should therefore remain fixed at least for very low eccentricities.
We remind the reader that if $\tilde{\Lambda}_{e}(x)>0$, then the annulus of the disc with
radius $x$ gains angular momentum.

For a fixed value of $h$, we see from Figure \ref{fig:torque_density}
that the region along which the torque is excited becomes more extended
with increasing eccentricity. 
On the other hand, in the upper panels ($h=0.07$) and middle panels ($h=0.04$), the peak of $\tilde{\Lambda}_{e}$ decreases with increasing eccentricity. 
For a given eccentricity, the value of $\tilde{\Lambda}_{e}$ at the peak decreases with increasing $h$. It scales as $h^{-4}$ for $e=0$, and as $h^{-5/2}$ for $e\simeq 0.08$.

Although $\tilde{\Lambda}_{e}$ is strictly antisymmetric 
[i.e. $\tilde{\Lambda}_{e}(-x)=-\tilde{\Lambda}_{e}(x)$] only for $e=0$, it holds that $\tilde{\Lambda}_{e}\gtrsim 0$ 
at $x\gtrsim 0$ and $\tilde{\Lambda}_{e}\lesssim 0$ at $x\lesssim 0$, except for the curves
corresponding to $h=0.025$ and $e>0.12$.
In particular, for $h=0.025$ and $e=0.16$, rings at $x\simeq 0.2$ lose 
angular momentum. This happens
at distances slightly larger than apocentre, where the planet
feels a tail wind \citep[e.g.,][]{cre07,bit10,mut11,san19}. 

It is likely that the 1D approach employed in the following sections
is not valid for those combinations of $h$ and $e$
for which $\tilde{\Lambda}_{e}(x)$ is far from being an antisymmetric function.
More precisely, we will consider only models for which $\tilde{\Lambda}_{e}(R)=0$ 
at a single radial distance $R_{\Lambda}$, or 
on an interval $[R_{\Lambda,1},R_{\Lambda,2}]$. In the latter case, we define $R_{\Lambda}$
as the mid-point of the interval $R_{\Lambda}=(R_{\Lambda,1}+R_{\Lambda,2})/2$.
In general, $R_{\Lambda}\simeq a$. For the models shown in Figure \ref{fig:torque_density},
the maximum difference between $R_{\Lambda}$ and $a$ occurs
for the model with $h=0.07$ and $e=0.16$, and it is $|R_{\Lambda}-a|=0.072a$.

Figure \ref{fig:tq_diff_Rsoft} compares $\tilde{\Lambda}_{e}(x)$ for different values 
of $\beta$, with $h$ and $e$ fixed to some characteristic values. 
We see that as $\beta$ increases, $\tilde{\Lambda}_{e}$ tends to decrease and flatten in the central parts of the excitation region. In Section \ref{sec:num_models}, $\beta$ will 
be calibrated 
using numerical simulations.
\subsection{One-sided torque}
\label{sec:one_sided}
It is useful to define $T_{1s}^{(i)}$ and $T_{1s}^{(o)}$ as the torque exerted by the planet on the disc interior to $R=a$ and exterior to $R=a$, respectively. More precisely,
these inner and outer one-sided torques can be computed from $\Lambda_{e}$ as:
\begin{equation}
T_{1s}^{(i)}= \int_{0}^{a} \Lambda_{e} dR,
\end{equation}
and
\begin{equation}
T_{1s}^{(o)}= \int_{a}^{\infty}\Lambda_{e} dR.
\end{equation}
If the torque is mainly excited at the bottom of the gap, where the surface density
is $\Sigma_{\rm gap}$, we can write:
\begin{equation}
T_{1s}^{(i)}\simeq q^{2} \Sigma_{\rm gap} \omega^{2} a^{3} \int_{0}^{a} 
\tilde{\Lambda}_{e}(R)\, dR,
\end{equation}
and
\begin{equation}
T_{1s}^{(o)}\simeq q^{2} \Sigma_{\rm gap} \omega^{2} a^{3} \int_{a}^{\infty} 
\tilde{\Lambda}_{e} (R)\,dR.
\end{equation}

The computation of the one-sided torques involves a double integral
(over $\xi$ to compute $\tilde{\Lambda}_{e}$ in Eq. (\ref{eq:torque_density0main}), and 
then over $R$). However, we show 
that the integral over $R$ can be performed analytically in Appendix \ref{sec:one_sided_tq_app}.

If instead of defining the separation between inner and outer disc at $R=a$, 
it is defined at $R=R_{\Lambda}$,  we would obtain slightly larger values for
$|T_{1s}^{(i)}|$ and $T_{1s}^{(o)}$ but the difference is small. For instance,
for the case $h=0.07$ and $e=0.16$, where the $|R_{\Lambda}-a|=0.072a$ (see
Section \ref{sec:tq_density}), the torques increase by $8$ percent.

The magnitude of the inner one-sided torque $|T_{1s}^{(i)}|$ is identical
to  $T_{1s}^{(o)}$ only for $e=0$. In general, the magnitudes of the inner
and outer one-sided torques are slightly different. 
We therefore introduce the auxiliary parameter $\lambda$, 
which measures how much $|T_{1s}^{(o)}|$ differs
from $|T_{1s}^{(i)}|$:
\begin{equation}
\lambda \equiv\frac{ T_{1s}^{(o)}+T_{1s}^{(i)}}{|T_{1s}^{(o)}|+|T_{1s}^{(i)}|}.
\end{equation}
If $T_{1s}^{(i)}\simeq -T_{1s}^{(o)}$ then $|\lambda|\ll 1$. On the other hand,
when $T_{1s}^{(o)}$ and $T_{1s}^{(i)}$ have the same sign, then $|\lambda|=1$.

Hereafter, we define the one-sided torque as
\begin{equation}
T_{1s}\equiv \frac{1}{2}(T_{1s}^{(o)}-T_{1s}^{(i)}). 
\end{equation}
This definition of $T_{1s}$ is physically relevant only when $|\lambda|\ll 1$.
Figure \ref{fig:tq_onesided} displays the dimensionless one-sided torque defined as
\begin{equation}
\tilde{T}_{1s}(h,e) \equiv \frac{T_{1s}}{f_{0}q^{2}\Sigma_{\rm gap} \omega^{2}a^{4} h^{-3}},
\label{eq:dimensionless_T1s}
\end{equation}
i.e. the torque relative to the circular case. The value of $\tilde{T}_{1s}=1$ at $e=0$
because we assumed $\beta=0.9$, as in Figure \ref{fig:torque_density}.
We see that $\tilde{T}_{1s}$ has a maximum at $e\simeq 0.7 h$, and then
it decreases monotonically. 

Since we are going to use the approach derived for
gaps opened by planets in circular orbits, for which $\lambda=0$, cases with relatively large values of
$|\lambda|$ (say $|\lambda|>0.2$) shall be discarded\footnote{Note, however, that
$\lambda$ itself does not determine if gaps are formed or not. In fact, gaps can form even if $|\lambda|\simeq 1$ \citep[see, e.g.,][]{iva15}.}.
In all cases shown in Figure \ref{fig:tq_onesided}, $|\lambda|\leq 0.05$,
except for those cases with $h=0.025$ and $e>0.125$. More specifically,
for $h=0.025$, $T_{1s}$ shows 
a sudden change in the slope at $e=0.125$. This steepening occurs when $|\lambda|\simeq 0.05$.
For $h=0.025$ and $e=0.15$ (red dot in the right panel of Figure \ref{fig:tq_onesided}), it holds that the inner and outer one-sided torques
are both negative.

We empirically find that $\tilde{T}_{1s}$ for $\beta=0.9$ can be well characterized by a fitting
function that depends on $\tilde{e}=e/h$:
\begin{equation}
\tilde{T}_{1s}(\tilde{e})= 
\left\{
\begin{array}{cl}
2-\left[\left(1+\frac{5}{4}\tilde{e}^3\right)^{-4}+\tilde{e}^2\right]^{1/4} &\textup{if  } \tilde{e}\leq2, \,\,\\
  &  \\
\frac{1}{2}\left(3\tilde{e}^{-0.43}-1\right) & \textup{if } \tilde{e} > 2.
\end{array}
\right.
\label{eq:fitT1s}
\end{equation}
The fitting error is less than $5$ percent, except for $\tilde{e}\geq 5$. 

\begin{figure}
\includegraphics[width=85mm, height=77mm]{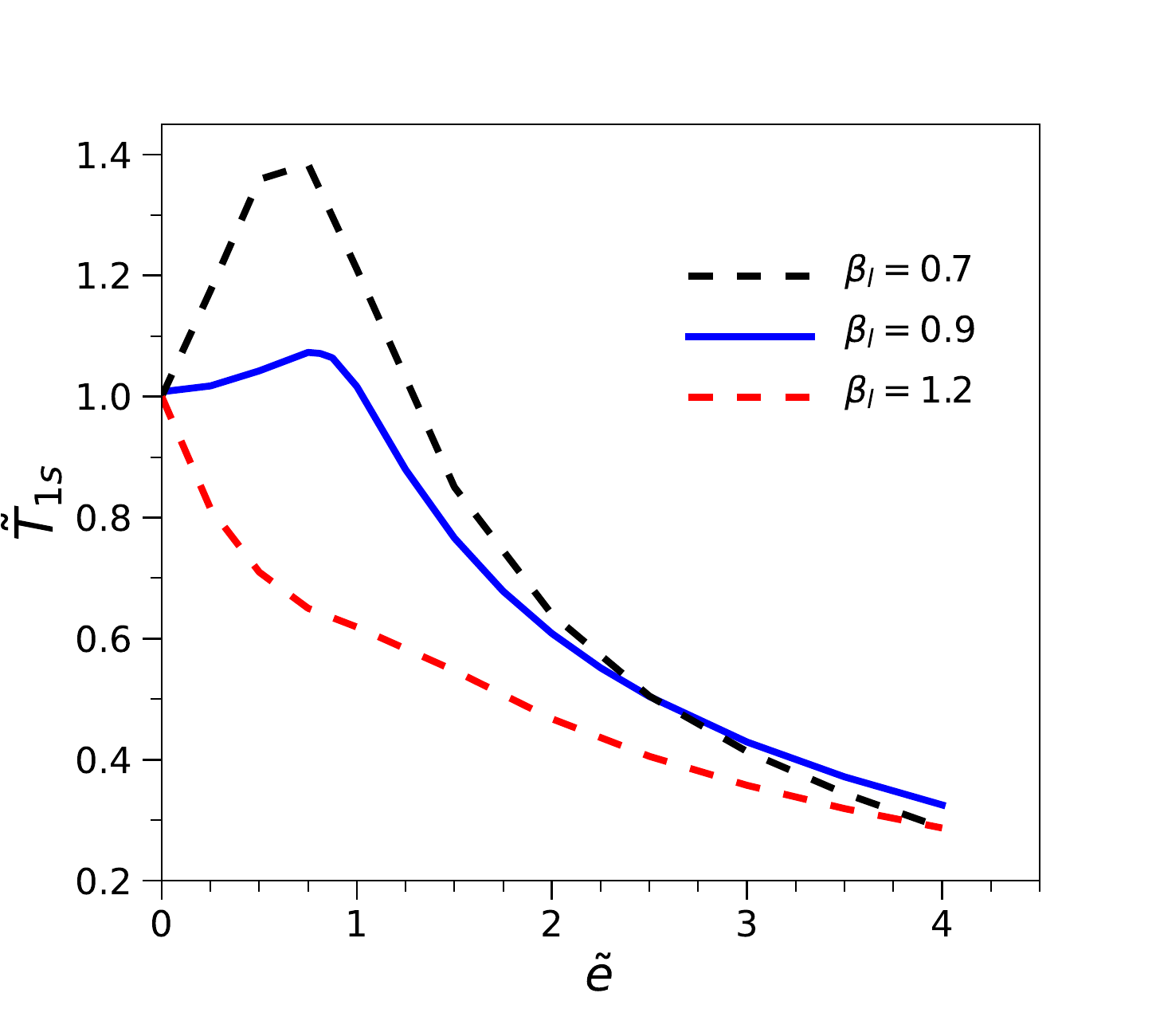}
\caption{Dimensionless one-sided torque $\tilde{T}_{1s}$ (relative to the circular case) calculated in the impulse approximation
as a function of $\tilde{e}=e/h$.  $\beta$ is given by Equation (\ref{eq:varying_beta})
with $\beta_{0}=0.9$. The values of $\beta_{l}$ are $0.7$ (dashed black curve),
$0.9$ (solid blue curve) and $1.2$ (dashed red curve).}
\label{fig:tq_varyingRsoft}
\end{figure}

Our reference value $\beta=0.9$ used so far was calibrated using the well-studied circular case
$\tilde{e}=0$ for which the planet is approximately corotating with the local gas disc.
However, as $\tilde{e}$ increases, the relative velocity of the planet with respect to the
gas also increases. If $\tilde{e}\gtrsim 2$, the motion of the planet is supersonic 
relative to the local gas velocity at any point of the orbit 
\citep[e.g.,][]{mut11,gri15}.
The transition between subsonic to supersonic motion could lead to a non-constant $\beta$. In other words,
we cannot rule out a slight dependence of $\beta$ on $\tilde{e}$.
To demonstrate the dependence
of the one-sided torque on $\beta$ varying with $\tilde{e}$,
Figure \ref{fig:tq_varyingRsoft} shows $\tilde{T}_{1s}$ assuming that
\begin{equation}
\beta (\tilde{e})= \beta_{l} +(\beta_{0} - \beta_{l}) \exp (-\tilde{e}),
\label{eq:varying_beta}
\end{equation}
for $\beta_{0}=0.9$ and two values of $\beta_{l}=0.7$ and $1.2$.
For the adopted form of Equation (\ref{eq:varying_beta}), $\beta=0.9$ when 
$\tilde{e}=0$ and $\beta\to\beta_{l}$ when the planet motion is supersonic at any orbital phase.
From Figure \ref{fig:tq_varyingRsoft}, we see that
$\tilde{T}_{1s}$ at $\tilde{e}\simeq 1$ in the model with $\beta_{l}=0.7$
is a factor of $2$ larger than in the model with $\beta_{l}=1.2$.
At values $\tilde{e}\gtrsim 2$, the value of $\tilde{T}_{1s}$ is considerably less sensitive to $\beta_{l}$.

The condition that the outer and inner torques should be similar in magnitude imposes a constraint on $\tilde{e}$. For instance, if we allow $\beta$ to vary in
the range $[0.7,1.2]$, then $|\lambda|$ is acceptably small  ($|\lambda|\leq 0.18$) provided that $\tilde{e}\leq 4$. 
Therefore, we will restrict our analysis of the gap depth to models having
$h\in [0.025,0.07]$ and $\tilde{e}\leq 4$.

\begin{figure}
\includegraphics[width=85mm, height=77mm]{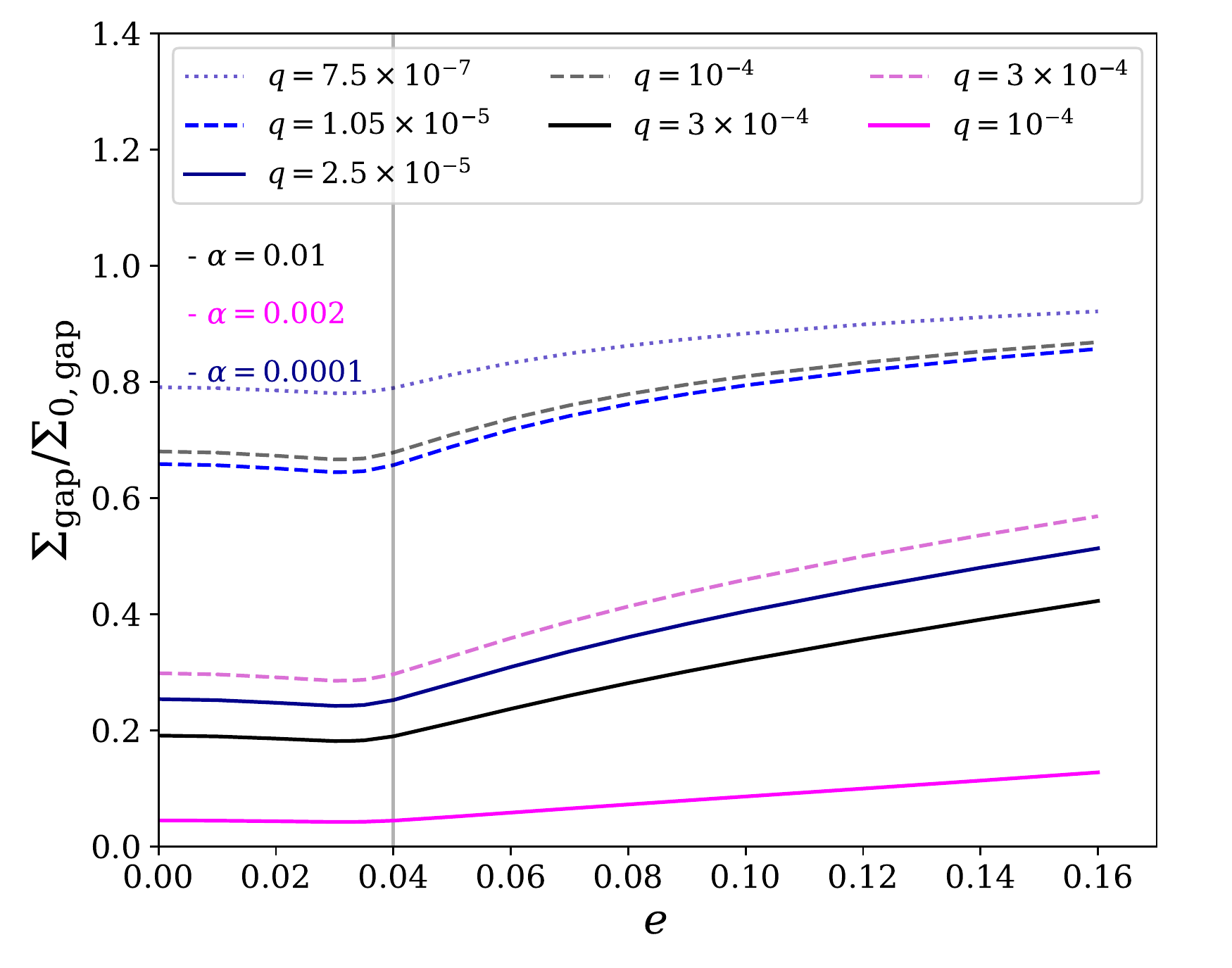}
\caption{Gap depth predicted in the zero-dimensional approximation as a function
of the eccentricity for different combinations of $q$ and $\alpha$. We assume
$h=0.04$ and $\beta=0.9$ in all cases.}
\label{fig:Sigma_gap_vs_ecc}
\end{figure}

\subsection{Gap depth in the zero-dimensional approach}

When the angular momentum is injected in a region that is narrower than 
the width of the gap, so that the one-sided torque deposited in the disc is $T_{1s}\propto
\Sigma_{\rm gap}$,
the dependence of the gap depth on $q$, $h$, $\alpha$ and $e$ 
can be derived in the so-called zero-dimensional approach as follows.

The one-sided torque $T_{1s}$
applied to the disc is $f_{0} \tilde{T}_{1s}q^{2}\Sigma_{\rm gap}\omega^{2}a^{4}h^{-3}$
(see Equation \ref{eq:dimensionless_T1s}).
On the other hand, the viscous flux of the angular momentum
is $-2\pi \nu \Sigma R^{3} d\Omega/dR\simeq 3\pi \nu \Sigma_{0,\rm gap} a^{2} \omega$
in a Keplerian disc.
Since the angular momentum deposited in the disc must be evacuated at the rate
of the viscous transport, one obtains
\begin{equation}
f_{0} \tilde{T}_{1s} q^{2} \Sigma_{\rm gap} \omega^{2}a^{4}h^{-3} \simeq 3\pi \nu 
(\Sigma_{0,\rm gap}-\Sigma_{\rm gap}) a^{2} \omega.
\label{eq:balance}
\end{equation}
We remind the reader that $\Sigma$ should be interpreted as the azimuthally averaged density. From Equation (\ref{eq:balance}), we obtain that 
the surface density at the bottom of the gap relative to the unperturbed density is
\begin{equation}
\frac{\Sigma_{\rm gap}}{\Sigma_{0,\rm gap}} \simeq \left(1+\frac{f_{0}\tilde{T}_{1s}q^{2} a^{2}\omega}
{3\pi \nu h^{3}}\right)^{-1}.
\label{eq:gap_depth_origin}
\end{equation}
In terms of the Shakura-Sunyaev $\alpha$ parameter, we write
\begin{equation}
\frac{\Sigma_{\rm gap}}{\Sigma_{0,\rm gap}} \simeq \left(1+\frac{f_{0} \tilde{T}_{1s}q^{2} }
{3\pi \alpha h^{5}}\right)^{-1} = \left( 1+ \frac{f_{0} K_{e}}{3\pi}\right)^{-1},
\label{eq:gap_depth_obvious}
\end{equation}
where $K_{e}=  \tilde{T}_{1s} q^{2}/(\alpha h^{5})$. We will refer to Equation 
(\ref{eq:gap_depth_obvious})
as the zero-dimensional formula for the gap depth.
It is interesting to note that the depth predicted by these
formulas does not depend on the distance that the waves travel before the 
angular momentum they carry is deposited. For circular orbits, 
we recover the form for the depth of partial gaps carved by planets in circular orbit 
\citep[e.g.,][]{fun14,kan15,kan17,duf15a,gin18,dem20}, because
$\tilde{T}_{1s}=1$, and thus $K_{e=0}\equiv K_{0}= q^{2}/(\alpha h^{5})$.

Figure \ref{fig:Sigma_gap_vs_ecc} shows how 
$\Sigma_{\rm gap}/\Sigma_{0,\rm gap}$ depends 
on the eccentricity for $h=0.04$ in our theoretical model (with $\beta=0.9$).
We see that $\Sigma_{\rm gap}/\Sigma_{0,\rm gap}$ exhibits a slight decrease 
(but remains nearly constant) at $e\leq h$, whereas it slowly increases (the gap becomes shallower) for larger values of $e$.

\section{Comparison with numerical simulations}
\label{sec:num_models}

\subsection{Description of the numerical simulations}

We perform 2D hydrodynamical simulations of the evolution of the gas surface density in a protoplanetary disc harboring a planet on a fixed eccentric orbit. Our numerical model is basically the same as in \citet{Chametla2022} but we do not include dust particles. Our hydrodynamical simulations are set up as follows. We use the 2D FARGO-ADSG code \citep{Baruteau2008a,Baruteau2008b} with a locally isothermal equation of state  
\citep[as in the original FARGO code of][]{Masset2000}.
The gas pressure is defined by the ideal gas law. In all models,
our initial disc has a constant aspect ratio $h$ (with $h\in[0.025,0.07]$) and its surface
density obeys a simple power law $\Sigma_{0}=\Sigma_{0,a}(R/a)^{-1/2}$.
The radial domain extends from $0.5 a$ to $2.5 a$. We employ the damping boundary
conditions described in \citet[][]{deVal2006}. The computational mesh consists of
$384$ rings and $1206$ sectors in the radial and azimuthal directions, respectively. 
We consider a range of planet masses $q\in[7.5\times10^{-7},5\times10^{-4}]$.
As in \citet[][]{Chametla2022}, depending on the disc viscosity (here $\alpha\in[10^{-4},10^{-3}]$), we evolve the gas disc over 6000--4000 planet orbits to achieve a steady state (low-viscosity simulations require longer time spans).

Since our simulations are 2D, we model the (vertically averaged) gravitational potential 
of the planet by a softened Plummer model with the smoothing length $R_{\rm soft}$, 
which should be a sizable fraction of $H\equiv c_{s}/\Omega$ \citep[e.g.][]{mul12}.
In our fiducial simulations, we take 
$R_{\rm soft}={\mathcal E} H_{p}$, with ${\mathcal E}=0.4$.
When calculating the azimuthally averaged surface density, we excluded cells close
to the planet prior to the calculation to avoid contamination by the density accumulation in the central part of the Hill sphere.

\subsection{Gap depth in the zero-dimensional approximation: Comparison
with simulations}
\label{sec:sims_zeroth}

\begin{figure}
\includegraphics[width=89mm, height=166mm]{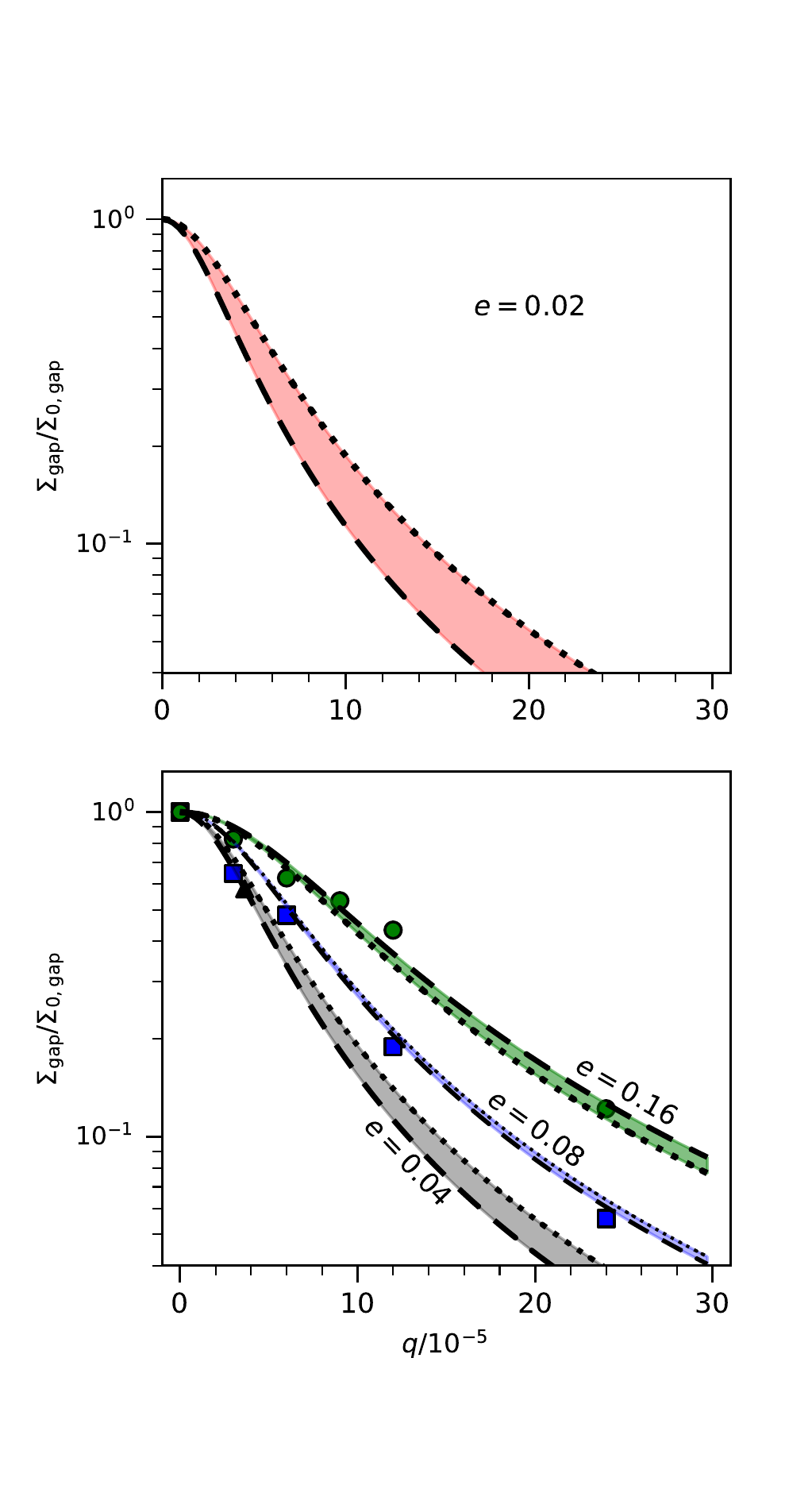}
\vskip -1.0cm
\caption{Gap depth as a function of $q$ in a disc with $h=0.04$ and $\alpha=10^{-3}$.
The shaded regions indicate the gap depth in the zero-dimensional approach 
when $\beta$ is within the interval $[0.7,0.9]$. Dashed lines correspond to $\beta=0.7$ and
dotted lines to $\beta=0.9$.
 Symbols correspond to the results of 
simulations using the fiducial parameters (${\mathcal E}=0.4$, $h=0.04$ and
$\alpha=10^{-3}$). Circles, squares and triangles correspond to $e=0.16$, $0.08$
and $0.04$, respectively.
}
\label{fig:gap_depth_16sims}
\end{figure}

The range of validity of Equation (\ref{eq:gap_depth_obvious}) has been extensively
studied for the case of circular orbits. By comparing
with simulations, \citet{kan17} find that Equation (\ref{eq:gap_depth_obvious}) 
correctly reproduces $\Sigma_{\rm gap}$ provided that 
$K_{0}=q^{2}/(\alpha h^{5})\leq 10^{3}$.
For larger values of $K_{0}$, Equation (\ref{eq:gap_depth_obvious}) 
overestimates $\Sigma_{\rm gap}$. However, even for $K_{0}<10^{4}$,
Equation (\ref{eq:gap_depth_obvious}) correctly predicts the averaged surface
density in an annulus with inner and
outer radii $a-\delta$ and $a+\delta$, respectively, 
where $\delta\equiv 2{\rm max}(R_{H},H)$
and $R_{H}$ is the Hill radius \citep[see also][]{fun14}.
In the following, we explore whether Equation (\ref{eq:gap_depth_obvious})
for eccentric orbits remains consistent with simulations in which we measure
the surface density at the gap bottom as $\Sigma_{\rm gap}=\min(\Sigma(R))$
(after removing the density peak around the planet
as in \citet{duf15b} and \citet{duf20}).

As said in Section \ref{sec:one_sided}, $\beta(\tilde{e})$ is an unknown function
with the constraint $\beta(0)\simeq 0.9$. The sensitivity of the gap depth on $\beta$ 
in the zero-dimensional
approximation is illustrated in Figure \ref{fig:gap_depth_16sims}. This figure shows
the predicted gap depth as a function of $q$. The coloured bands
indicate the range of gap depths when $\beta$ is varied within the interval $[0.7,0.9]$.
Since $R_{\rm cut}=\beta H= \beta R_{\rm soft}/ {\mathcal E}$,
the limit values of $\beta$ imply
that $R_{\rm cut}= 1.75R_{\rm soft}$ and $2.25R_{\rm soft}$, respectively\footnote{In
the limit of a supersonic body in rectilinear orbit, the relationship between
$R_{\rm cut}$ and $R_{\rm soft}$ is $R_{\rm cut}\simeq 4R_{\rm soft}/3$ 
\citep{mut11}.}. 
The thickness of the shaded region is relatively large for $e=0.02$. 
However, for $0.04\leq e\leq 0.16$,
the gap depth is quite insensitive to variations of $\beta$ between $0.7$ and $0.9$.

In the lower panel of Figure \ref{fig:gap_depth_16sims}, the gap depths as found
in the simulations have been also included.
The comparison between our theoretical models and simulations
of {\it partial gaps}
suggests that if we fix the softening-to-thickness ratio to ${\mathcal E}=0.4$,
then the parameter $\beta$ should be taken between $0.7$ to $0.9$ to reproduce
the gap depth for $e>0.04$ (i.e. $\tilde{e}\geq 1$).
Note that the precise value of $\beta$ depends on the adopted value for ${\mathcal E}$
because if we adopt a large value for ${\mathcal{E}}$, the depth of the gravitational 
potential created by 
the planet becomes less profound and therefore a shallower gap is formed.

\begin{figure}
\includegraphics[width=87mm, height=78mm]{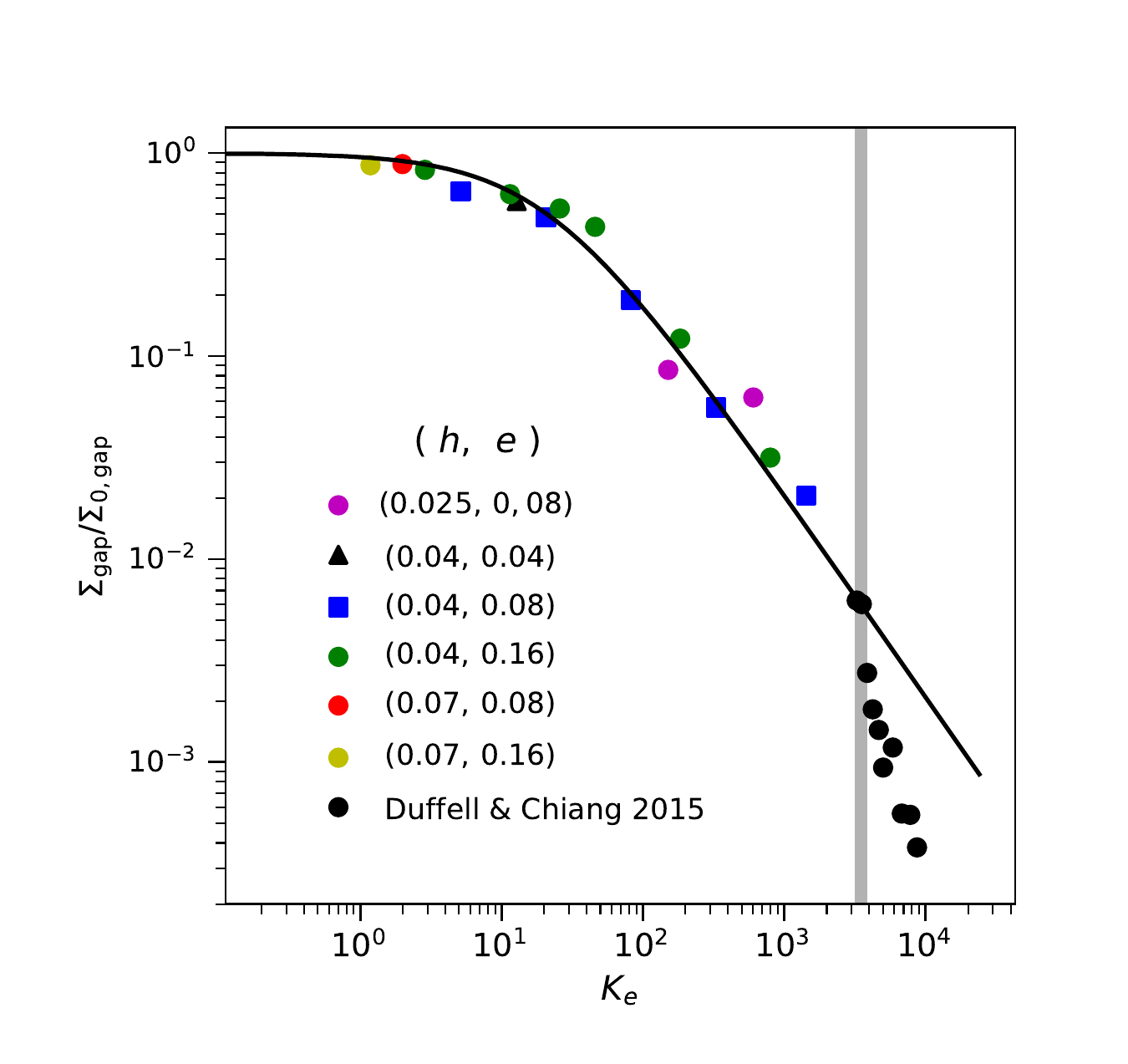}
\caption{Gap depth as a function of $K_{e}\equiv \tilde{T}_{1s} q^{2}/(\alpha h^{5})$ for different planet-disc models.
$\tilde{T}_{1s}$ was computed using the analytical fit given in Eq. (\ref{eq:fitT1s}).
All cases assume $\alpha=10^{-3}$,
except for the simulations of \citet{duf15b} who use $\alpha=2\times 10^{-3}$.
The solid curve represents Equation (\ref{eq:gap_depth_obvious}). The vertical grey line
indicates the value of $K_{e}=3.5\times 10^{3}$ for which Equation (\ref{eq:gap_depth_obvious}) begins to deviate 
from the values found in simulations.}
\label{fig:depth_Ke}
\end{figure}

\begin{figure}
\includegraphics[width=90mm, height=226mm]{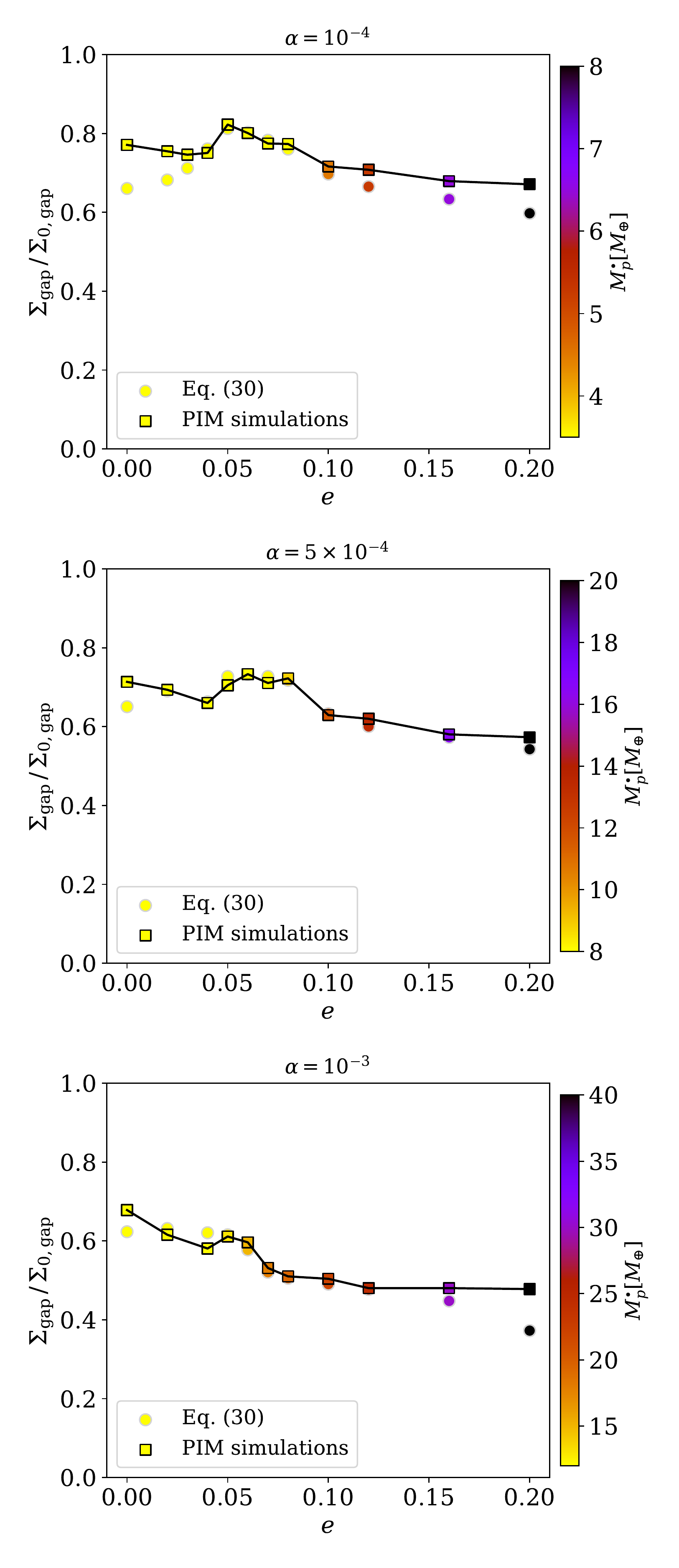}
\vskip -0.3cm
\caption{Comparison between the depth of the gap in numerical simulations and
model predictions (Equation \ref{eq:gap_depth_obvious}) in the low $\alpha-$viscosity regime ($\alpha\leq10^{-3}$).
The mass of the planets corresponds to $M_{p}^{\bullet}$, their PIM in Earth masses.}
\label{fig:PIM}
\end{figure}

Figure \ref{fig:depth_Ke} shows the gap depth as a function of $K_{e}$, as found in
our numerical simulations with
$q\in[3\times10^{-5},5\times10^{-4}]$, $h\in[0.025,0.07]$, $e\geq 0.04$, 
and $\alpha=10^{-3}$, but with the constraint $\tilde{e}\leq 4$.
For each planet-disc model, $\tilde{T}_{1s}$ was
calculated using the analytical fit given in Equation (\ref{eq:fitT1s}).
We also included the gap depth reported in \citet{duf15b}, who also carry out
numerical simulations of the gap formed by eccentric planets. In their models, 
$q=10^{-3}$, $h=0.036$, $e\leq 0.12$ and $\alpha=2\times 10^{-3}$. 

Simulations in Figure \ref{fig:depth_Ke} are compared with predictions
given by Equation (\ref{eq:gap_depth_obvious}).
We see that our model predicts the correct scaling of the gap depth up to
$K_{e}\simeq 3\times 10^{3}$. In terms of the gap depth, the scaling deviates
from simulation results when $\Sigma_{\rm gap}/\Sigma_{0,\rm gap}\lesssim 10^{-2}$.
For $K_{e}>3\times 10^{3}$, $\Sigma_{\rm gap}$ in the simulations are below the
values obtained from Equation (\ref{eq:gap_depth_obvious}).
We recall that for planets in circular orbits, 
\citet{kan17}  also find that Equation (\ref{eq:gap_depth_obvious}) predicts
$\Sigma_{\rm gap}$ correctly only when the gap is not very deep ($K_{0}\leq 10^{3}$).

Even in the case of planets on circular orbits, the values of $\Sigma_{\rm gap}/\Sigma_{0,\rm gap}$ obtained from simulations exhibit
some scatter around the values predicted by 
Equations (\ref{eq:fitT1s}) and  (\ref{eq:gap_depth_obvious}) \citep[e.g.][]{duf15a,kan17}.
Since the gas flow within the gaps becomes more complex for eccentric planets, we would expect a larger scatter in 
the $\Sigma_{\rm gap}$-$K_{e}$ relationship when planets with $e>0$
are considered.
The increasing complexity of
the flow is also reflected in the fact that, in general, the minimum of the surface
density in the gap does not occur exactly at $R=a$ \citep[e.g., see Fig. 7 in][for $e=0.2$]{Chametla2022}.
However, although the number of simulations in Figure \ref{fig:depth_Ke} is too small 
to quantify the scatter, it seems that it does not exceed the scatter found for circular orbits.

\begin{figure}
\includegraphics[width=86mm, height=120mm]{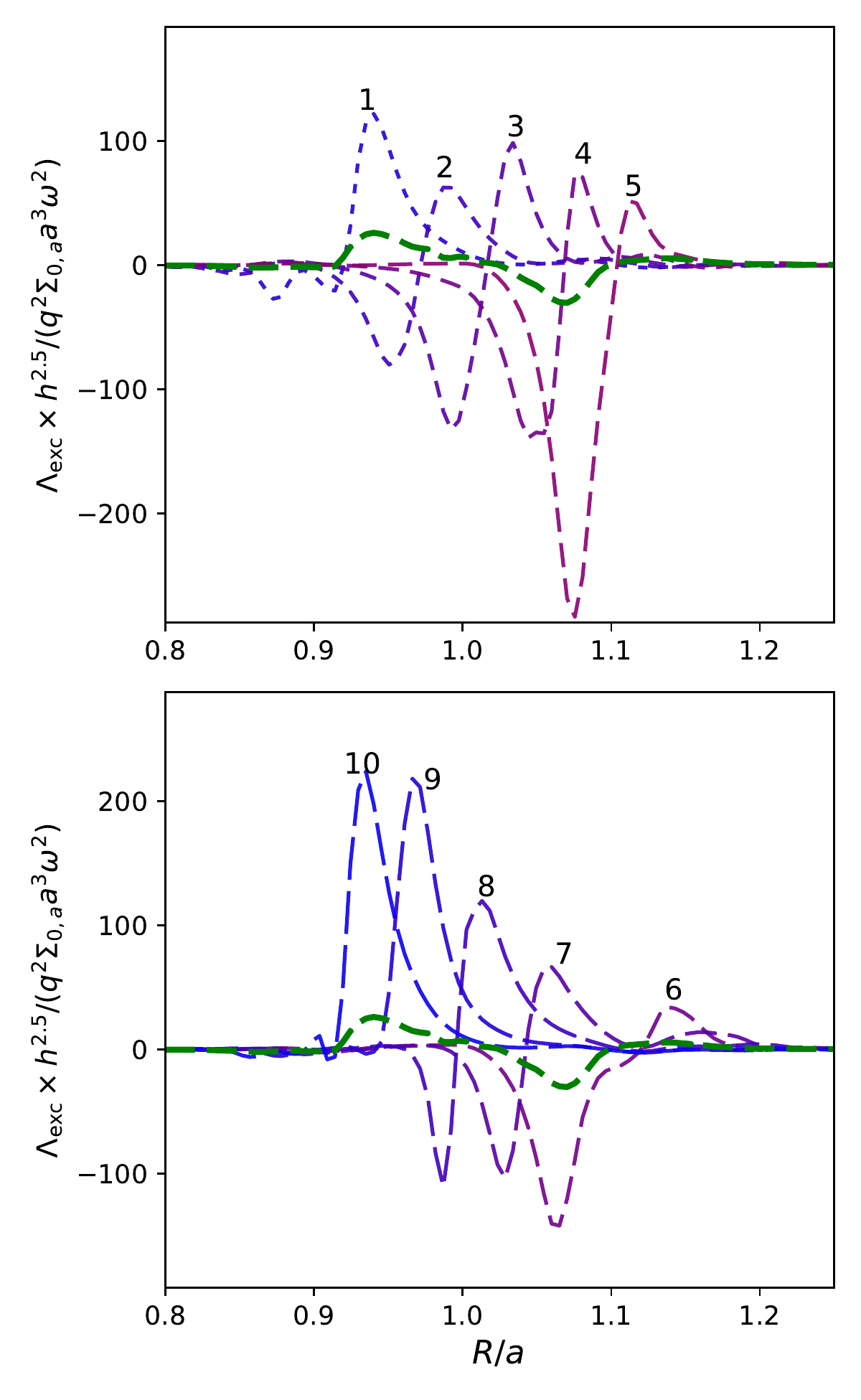}
\caption{Density of the excitation torque exerted on the disc by the planet, taken from a simulation with $q=6\times 10^{-5}$, $h=0.04$, $\alpha=10^{-3}$ and $e=0.08$,
at ten orbital phases in the last orbit (blue shaded lines). In the top panel, the planet
is moving from pericentre to apocentre. In the bottom panel, it moves from apocentre
to pericentre.
The difference between consecutive times is $0.2 \pi/\Omega$. 
The labels indicate the time ordering. The curve labelled $5$ is the
torque density when planet is at apocentre. The curve labelled $10$ corresponds to 
pericentre.
The dashed green curve represents the 
torque density averaged over one orbit.
 }
\label{fig:tq_density_sims_ecc008}
\end{figure}

We also have a set of simulations that was initially designed to determine the PIM,
denoted by $M_{p}^{\bullet}$, for planets
in eccentric orbit \citep{Chametla2022}. This set of simulations can be useful to 
check the accuracy of the gap depth scaling in the shallow gap regime, because the
deepest gap has $\Sigma_{\rm gap}/\Sigma_{0,\rm gap}=0.48$.
All these simulations use $h=0.04$ and $\mathcal{E}=0.4$.

We compare the gap depth using our scaling relation
[Equations (\ref{eq:fitT1s}) and (\ref{eq:gap_depth_obvious})], with
the results of hydrodynamical simulations 
in Figure \ref{fig:PIM}.
The colour bar indicates the corresponding 
$M_{p}^{\bullet}$. As expected, it increases with increasing the eccentricity.
We find that the depth of the gap shows
remarkable agreement with those of the simulations, except at the end points,
i.e. at $e=0$ and $e=0.2$.
We have included the value $e=0.2$ (which has $\tilde{e}=5$) 
for completeness, but we already anticipated in Section \ref{sec:one_sided} that the range
of validity of our model is $\tilde{e}\leq 4$. Thus, the failure of the
model at $\tilde{e}\simeq 5$ was to be expected.

At $e=0$, our scaling relation predicts deeper gaps than those found in the
simulations, especially for $\alpha=10^{-4}$. For zero eccentricity, our model is
not different to the model in \citet{duf15a}, who
already noticed some deviations in the circular case when the parameter $K_{0}$
is larger than $10$. In Figure \ref{fig:PIM},
the models where the planet is on a circular orbit ($e=0$) have $K_{0}=10.5-13.5$.
For $K_{0}=10$, \citet{duf15a} finds $\Sigma_{\rm gap}/\Sigma_{0,\rm gap}=0.75$ in
his simulations with $h=0.025$;  a shallower gap than the predicted value ($0.65$).
However, he did not explore cases with $\alpha<0.005$. Here we find that if the
$\alpha$-parameter is very low $\alpha \lesssim 10^{-4}$, some deviations 
also occur for $K_{0}\simeq 10$.

At $e\simeq h$, the scaling relation is consistent with the gap depth
for the three values of the $\alpha$ viscosity parameter. In fact,
it can reproduce the bump at $e\approx 0.05-0.07$ 
in the diagrams of Figure \ref{fig:PIM}. 

For $\alpha\geq 5\times10^{-4}$, the scaling relation
satisfactorily predicts the gap depth (with an error less than $10\%$ as long as 
$e\leq 0.16$ or, equivalently, $\tilde{e}\leq 4$).

\begin{figure*}
\includegraphics[width=200mm, height=144mm]{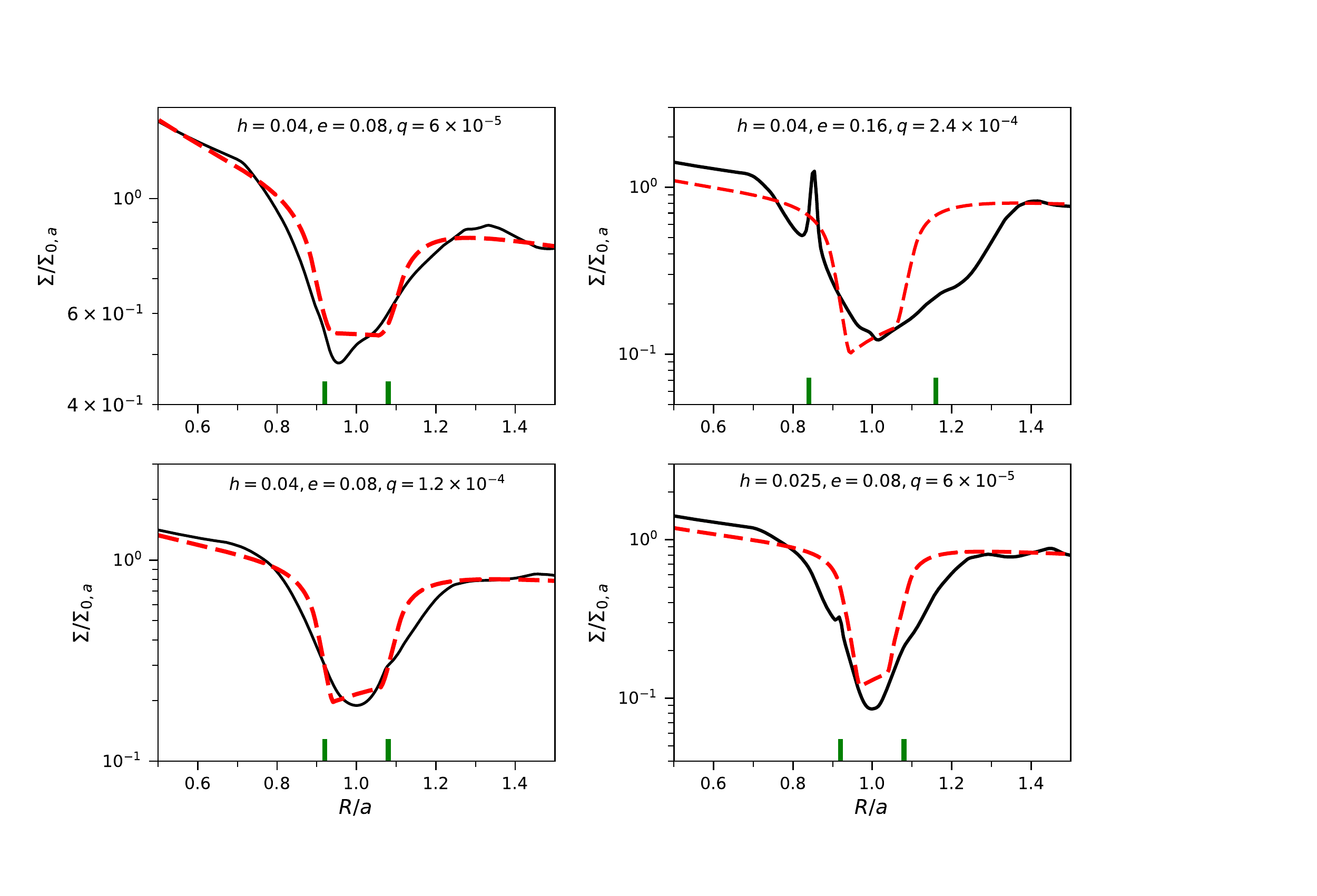}
\vskip -0.5cm
\caption{Comparison between the radial variation of the surface density in the simulations (solid lines)
and in model B assuming that the one-sided torques are excited in a narrow region around $R=a$ (dashed red lines). We see that the gaps in dashed red lines
have the same depth as the gaps in the simulations. All the simulations use $\alpha=10^{-3}$.
The green vertical lines mark the pericentre and apocentre of the planet. }
\label{fig:shallow_deep_tq}
\end{figure*}

\subsection{On the limitations of the impulse approximation}

The impulse approximation, or so-called dynamical friction approach,
only takes into account
the tidal (local) deformation of the disc flow when a gas parcel encounters the planet.
\citet{san19} numerically shows that this 
non-resonant local approach predicts the total torque acting on an eccentric planet correctly,
provided that $e/h \gtrsim 2.5$ and the softening radius is low enough.

The instantaneous excitation torque density $\Lambda_{\rm exc}(t,R)$ can be extracted 
from our simulations. Figure \ref{fig:tq_density_sims_ecc008} shows the excitation 
torque density at $10$ orbital phases during the last orbit of the run 
$q=6\times 10^{-5}$, $h=0.04$, $\alpha=10^{-3}$, and $e=0.08$. 
The torque density is negative at almost all radii when the planet is at apocentre 
(curve labelled 5),
because it moves more slowly than the local gas. At pericentre (curve labelled
10), the planet rotates faster and it pushes the gas leading to a positive torque
\citep[see also][]{bit10}. Individually, the shape and amplitude of each curve in 
Figure \ref{fig:tq_density_sims_ecc008} is consistent with what we expect in the
impulse approximation.

However, the excitation torque density after orbital averaging is positive in the inner
disc ($R<a$) and negative in the outer disc ($R>a$). 
This implies that, if the deposition of angular momentum were local, an
overdense ring rather than a gap would form.
This example makes clear the importance of wave propagation to set a criterion for gap formation in the case of planets in eccentric orbit\footnote{In the context of the
gravitational interaction of an eccentric binary system with the 
circumbinary disc, it is assumed that a gap is opened at the
$(m,n)$ Lindblad resonance if the torque $|T_{m,n}|$ (where $m$ is the azimuthal wave in the disc and
$n$ is the time harmonic number) is larger than the viscous
torque \citep{art94,mir15}. }. 
Our model based on the impulse 
approximation cannot successfully capture the excitation torque density, 
because the calculation involves a
delicate orbital average.
It would be interesting to see if models based on the propagation of density waves
\citep[e.g.,][]{pap00} are able to reproduce $\Lambda_{\rm exc}(t,R)$ as obtained in
the simulations.

In this paper, we used the impulse approximation as a framework to derive 
simple scalings that we calibrated against hydrodynamical simulations. Thus, our 
derivation of the scaling of the gap depth is ultimately an empirical relation. 
The fact that we could account
for the gap depth scaling by simply assuming a constant value of $\beta$ may be coincidental.

\begin{figure*}
\includegraphics[width=180mm, height=144mm]{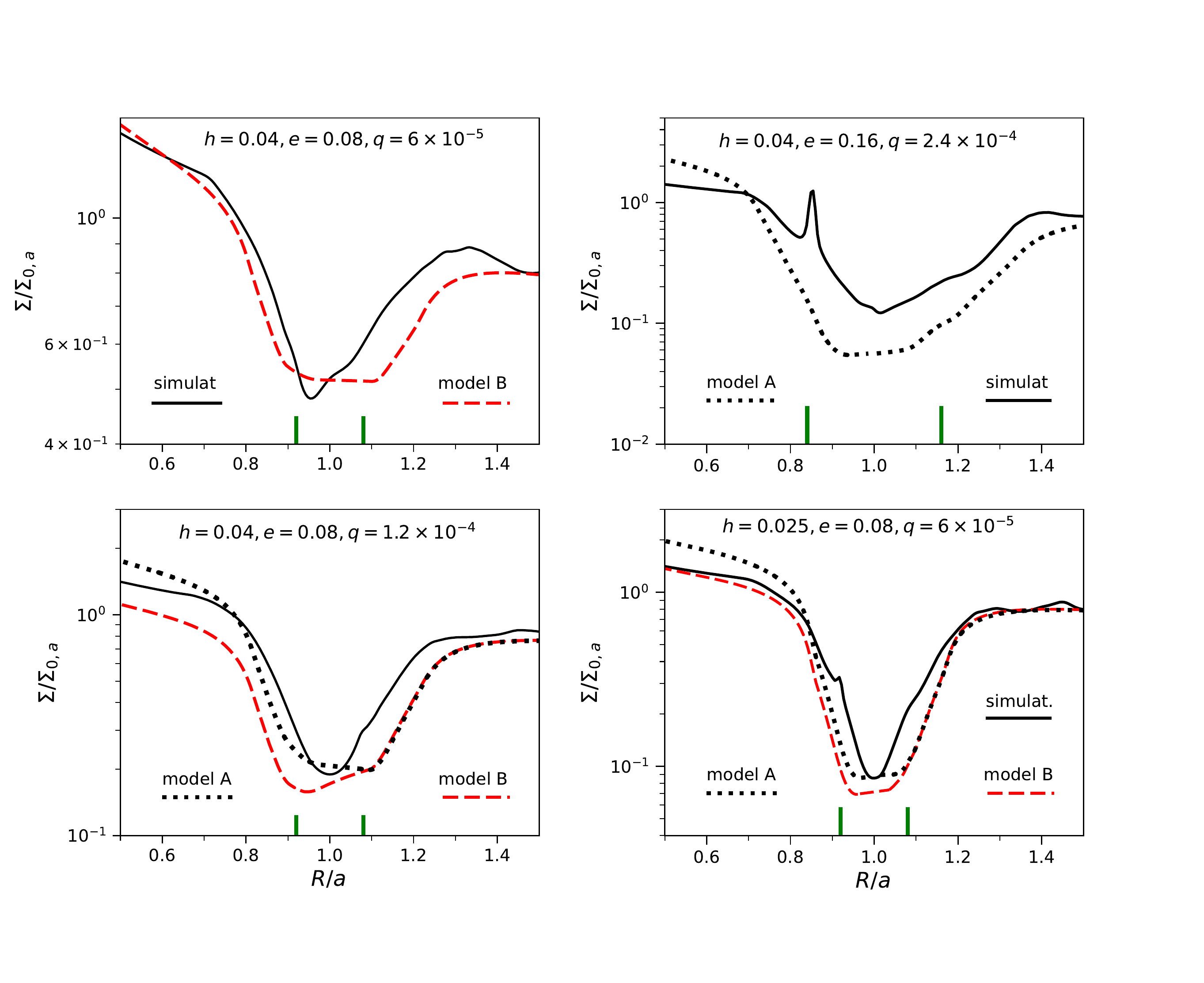}
\caption{Radial profiles of gaps 
taken from our hydrodynamical simulations (solid black curves),
together with the predicted profile assuming non-local deposition of angular momentum
in model B (dashed curves). In the middle panel, we also show the profile for model A
(dotted line).
In all models we take $\beta=0.9$. In models A, we used $\nu=10^{-3}h^{2} a^{2}\omega$.
In the simulations and in models B, we used $\alpha=10^{-3}$. 
The green vertical lines mark the pericentre and apocentre of the planet.
 }
\label{fig:shallow_deep_nonlocal}
\end{figure*}

\subsection{The shape of the gap}
\label{sec:shape_gap}
In this section we explore if simple models for density wave excitation and damping
can describe the shape of a gap around an eccentric planet.
Equation (\ref{eq:Sigma_time_1}) describes how the gap is opened when we insert a planet in the disc.
We will use the dimensionless variable $b\equiv \sqrt{R/a}$, and
parameterize $\nu= \nu_{0} a^{2}\omega f(b)$, where $\nu_{0}$ is the
dimensinonless kinematic viscosity at $b=1$. In a disc with a constant 
kinematic viscosity, we have $f(b)=1$.  In a disc with a constant 
Shakura-Sunyaev $\alpha$ viscosity parameter \citep{sha73}, $\nu_{0}=\alpha h^{2}$
and $f(b)=b$.
In terms of the dimensionless variables $b$ and $\tau\equiv 3\nu_{0}\omega t/4$, 
Equation (\ref{eq:Sigma_time_1}) can be recast as
\begin{equation}
\frac{\partial \Sigma}{\partial \tau} = \frac{1}{b^{3}} \frac{\partial}{\partial b}
\left(\frac{\partial}{\partial b}
\left(b f(b) \Sigma\right) -\frac{2b\Lambda_{\rm dep}}{3\pi \nu_{0} \omega^{2} a^{3}}\right),
\label{eq:dSigma_dtau}
\end{equation}
where we used $\Omega = \omega/b^{3}$ for a thin Keplerian disc.

Let us denote $\Sigma_{0}(b)$ the surface density in a steady state before the planet is
introduced in the disc. Equation (\ref{eq:dSigma_dtau}) with $\Lambda_{\rm dep}=0$
leads to
\begin{equation}
\frac{d}{db} \left(b f(b) \tilde{\Sigma}_{0}\right)=F_{0},
\end{equation}
where $\tilde{\Sigma}_{0}(b)\equiv \Sigma_{0}/\Sigma_{0}(a)=\Sigma_{0}/\Sigma_{0,a}$ is the dimensionless surface density $F_{0}$ is a constant of integration. As explained in \citet{iva15}, 
there are two special cases of interest: $F_{0}=0$ represents a zero mass flux along the disc,
whereas $F_{0}=1$ describes a disc with a zero angular momentum flux.
For instance, for a disc with $F_{0}=0$ and constant viscosity ($f(b)=1$), we get
$\tilde{\Sigma}_{0}=(R/a)^{-1/2}$. The same radial dependence occurs in a steady-state
disc with $F_{0}=1$, $\alpha=\mathrm{const}$ and $h=\mathrm{const}$.
For shortness, we will refer to models with $F_{0}=0$ and $\nu=\mathrm{const}$
as models A, whereas models B assume $F_{0}=1$, $\alpha=\mathrm{const}$
and $h=\mathrm{const}$.

In the presence of the planet, the steady-state (dimensionless) surface density satisfies:
\begin{equation}
\frac{d}{db}(b f(b) \tilde{\Sigma}) -\frac{2b\Lambda_{\rm dep}}{3\pi \nu_{0}\Sigma_{0,a} \omega^{2}a^{3}}=F_{0}.
\label{eq:nonlocal_deposition}
\end{equation}
For $F_{0}=0$ and $f(b)= {\rm const}$, the solutions of Equation (\ref{eq:nonlocal_deposition}) are positive ($\tilde{\Sigma}>0$ at $b>0$).
There exist situations where Equation (\ref{eq:nonlocal_deposition}) has a solution with unrealistic negative values of the surface density.
Appendix \ref{sec:analytical_diff_eq} presents an analytical example showing that  
these unphysical solutions may occur for $F_{0}=1$, 
as long as the viscosity is so low that the inward advective transport of angular momentum is not sufficient to counterbalance 
the rate of angular momentum deposited by the planet.

For planets in circular orbits, the deposition torque density is approximately given by
\begin{equation}
\Lambda_{\rm dep}(R)= T_{1s} \,\bigg|\frac{d\Phi}{dR}\bigg| \,{\rm sign} (R-a)
\end{equation}
where 
\begin{equation}
\Phi(R)=
\left\{
\begin{array}{cl}
1 &\textup{if  }  \,\,\tau(R)<\tau_{\rm sh} \\
  &  \\
\sqrt{\tau_{\rm sh}/\tau} &\textup{if  }  \,\,  \tau(R)>\tau_{\rm sh} \\
&\\
0 & \textup{otherwise,}
\end{array}
\right.
\label{eq:damping_Rafikovcc}
\end{equation}
where $\tau_{\rm sh}$ is the dimensionless wave-to-shock timescale (see equations 35 in \citet{raf02}
for the non-dimensionalisation used), given by 
\begin{equation}
\tau_{\rm sh}=1.89+0.53 h^{3}/q,
\label{eq:shock_position}
\end{equation}
and
\begin{equation}
\tau(R) = \frac{3}{(2 h^{2})^{5/4}} \bigg|\int_{1}^{R/a} |s^{3/2}-1|^{3/2}
s^{(5p_{2}+p_{1})/2-11/4} ds \bigg|
\label{eq:rafikov43cc}
\end{equation}
where $p_{1}$ and $p_{2}$ represent the power index of
the surface density and sound speed of the unperturbed disc, respectively 
($\Sigma\propto R^{-p_{1}}$ and
$c_{s}\propto R^{-p_{2}}$). In our case, $p_{1}=p_{2}=1/2$. 
We need to mention that Equation (\ref{eq:damping_Rafikovcc}) is a fit to the solution calculated by \citet{goo01} and \citet{raf02} \citep[see also][]{duf15a}.

We computed the gap shape using Equations (\ref{eq:nonlocal_deposition})--(\ref{eq:rafikov43cc}), 
assuming that the one-sided torques have magnitude $f_{0}\tilde{T}_{1s}q^{2}\Sigma_{\rm gap}
\omega^{2}a^{4}h^{-3}$, with $\tilde{T}_{1s}$ given by Equation (\ref{eq:fitT1s}).
Therefore, it is implicitly assumed that the torques
are excited in a very small region around $R=a$.
The equations are solved using an iterative scheme \citep{gel14}. 
We impose the unperturbed surface density $\Sigma_{0}(R_{\rm max})$ at a certain large external boundary radius $R_{\rm max}$. We use $R_{\rm max}=4a$.

Figure \ref{fig:shallow_deep_tq} shows variation of the gap profile with 
$q$, $h$ and $e$ together with the resultant theoretical profile in model B.
As expected from the success of the zero-dimensional
formula, the theoretical gaps have the same depths as in the simulations.
However, the predicted gaps are generally more narrow, especially in the case 
with $e=0.16$.
Interestingly, for shallow gaps with $\Sigma_{\rm gap}/\Sigma_{0,\rm gap}>0.4$, this approximated model predicts the width of the gap reasonably well.

A natural route to have wider gaps is to relax the assumption that the torque is injected
in a small region around $R=a$ but in a more extended region.  However, the excitation
region cannot be arbitrarily large because then the model could fail to reproduce 
the depth of the gap.
To illustrate this, we have made the following exercise.
We assume that (1) the excitation torque density is given by 
$\Lambda_{\rm exc}= \Sigma(R) \tilde{\Lambda}_{e}(R)$, where $\tilde{\Lambda}_{e}(R)$ 
as calculated in Section \ref{sec:tq_density} using $\beta=0.9$, 
and (2) the deposition torque at a radius $R$ is determined by the sum of the 
contribution of all the waves that damp at that radius, so that $\Lambda_{\rm dep}$ is
the convolution of $\Lambda_{\rm exc}$ (see Appendix \ref{sec:appD} for
the equations we have solved).

Figure \ref{fig:shallow_deep_nonlocal} 
compares the predicted profiles with the numerical ones.
For clarity, the gap profile in model A is not plotted in the upper left panel
because the profiles in models A and B essentially overlap. 
In the upper right panel ($h=0.04$, $e=0.16$), we only show the results 
for model A because
the iterative method does not converge for model B. 
For the latter, we searched for solutions at subintervals
and found that there is no physical (non-negative) solution that
satisfies matching conditions in 
model B for these values of $h$ and $e$. 
For these parameters, it holds that
$\int_{R_{\Lambda}}^{\infty}\tilde{\Lambda}_{e}\,dR\simeq -1.05 \int_{0}^{R_{\Lambda}} \tilde{\Lambda}_{e}\,dR$.\footnote{We found that convergence to a physical solution can be recovered 
if we rescale $\tilde{\Lambda}_{e}$
by a factor $1.025$ at $R<R_{\Lambda}$ and by a factor $1/1.025$ at $R>R_{\Lambda}$
so that $\int_{0}^{R_{\Lambda}} \tilde{\Lambda}_{e}\,dR= -\int_{R_{\Lambda}}^{\infty}
\tilde{\Lambda}_{e}\,dR$.}

We see that this model systematically overestimates the width of the gaps. 
In particular, this model predicts an approximately flat region at the gap bottom 
which is not visible in the simulations.
On the other hand, for a disc with $h=0.04$ and a planet with $e=0.16$, model A 
predicts a surface density at the gap bottom three times smaller than measured in the 
simulation (upper right panel in Figure \ref{fig:shallow_deep_nonlocal}). 
In this case, the discrepancy between the
predicted gap profile and the simulations is already visible at $R=1.4a$.

More sophisticated models are required to describe the profile of gaps carved by
planets in eccentric orbit. In order to compute the deposition torque density,
it is necessary to calculate first the excitation torque density and then to build
a model to incorporate the wave propagation and dissipation. This is a challenging task.

\section{The level of eccentricity of planets embedded in protoplanetary discs}  

\label{sec:ecc_damping}

A variety of substructures such as spiral arms, rings and gaps have been discovered in many protoplanetary discs.
Some of the arms are complex, showing bifurcations, crossing points
or fragmentation. Some features of the spiral arms are difficult to explain if they are formed by a planet in circular orbit \citep[e.g.,][]{mon19,RZ2021} 
It has been suggested that the multiple spiral arms in
TW Hydrae, HD 34700A, AB Aurigae, and the pattern speeds of the 
two grand-design spirals in SAO 206462 could be explained by
the presence of a planet in eccentric orbit \citep{cal20,zhu22}.

The level of eccentricity of planets and embryos when they are still embedded in their 
natal disc is an issue of active research \cite[see][for a review]{paa22}.
In this paper, we have studied the gap opened by a single planet in eccentric orbit, 
with a simplifying assumption that its eccentricity remains constant. Our theoretical
model is intended to provide a basic estimate of the gap depth in a steady state.
In other words, 
it is implicitly assumed that the timescale for the eccentricity to change, $\tau_{e}\equiv
|e/\dot{e}|$, is longer than the gap opening timescale.

For single planets, the interaction with the protoplanetary disc dictates the rate of
damping or growth of the eccentricity. In linear theory, \citet{tan04}
demonstrated that for planets with $e\ll h$ embedded in a smooth laminar disc, 
\begin{equation}
\tau_{e<h} = 13 \left(\frac{q}{10^{-4}}\right)^{-1} \left(\frac{q_{\rm disc}}{10^{-3}}\right)^{-1}\left(\frac{h}{0.05}\right)^{4} {\rm orbits},
\end{equation}
where $q_{\rm disc}\equiv \Sigma a^{2}/M_{\star}$. For $e\geq 2h$, \citet{pap00}
suggest to rescale $\tau_{e<h}$ by $(e/h)^{3}$:
\begin{equation}
\tau_{e>2h}= 100 \left(\frac{e}{0.1}\right)^{3} \left(\frac{q}{10^{-4}}\right)^{-1} \left(\frac{q_{\rm disc}}{10^{-3}}\right)^{-1}\left(\frac{h}{0.05}\right) {\rm orbits}.
\label{eq:tau_ecc_2h}
\end{equation}

On the other hand, the gap opening timescale $\tau_{\rm gap}$ (for planets 
in circular orbit) is
\begin{equation}
\tau_{\rm gap}=240 \left(\frac{q}{10^{-4}}\right) 
\left(\frac{h}{0.05}\right)^{-7/2}\left(\frac{\alpha}{10^{-3}}\right)^{-3/2} {\rm orbits}
\end{equation}
\citep{kan20}.
If we demand that $\tau_{e>2h}\gtrsim 5\tau_{\rm gap}$, we find the condition
\begin{equation}
q_{\rm disc}\leq 8\times 10^{-5} \left(\frac{e}{0.1}\right)^{3} \left(\frac{q}{10^{-4}}\right)^{-2} \left(\frac{h}{0.05}\right)^{9/2} \left(\frac{\alpha}{10^{-3}}\right)^{3/2},
\label{eq:low-mass-disc}
\end{equation}
which is valid for planets that open shallow gaps. If we compare this result with
$q_{\rm disc}$ in the Minimum Mass Solar Nebula (MMSN):
\begin{equation}
q_{\rm disc}^{\rm (MMSN)} = 1.9\times 10^{-4} \left(\frac{R}{1\,{\rm AU}}\right)^{1/2}
\end{equation}
\citep{hay81}, we can say that condition (\ref{eq:low-mass-disc}) could be fulfilled  
at late stages of the lifetime of the disc and probably for a short period of time.
Condition (\ref{eq:low-mass-disc}) can be also met if the planet is in a cavity because 
the low value of the local disc surface density implies low values of $q_{\rm disc}$.

The classical assumption that the eccentricity of low-mass planets decays to zero
due to disc-planet interactions is correct only as long as thermal
effects are ignored.
The inclusion of radiative thermal diffusion and the energy release by low-mass accreting planets
may have also an impact on the time evolution of eccentricity.  It was shown that if the luminosity of a low-mass
planet is large enough, the eccentricity grows to values comparable to $h$ \citep{ekl17,chr17,fro19,rom22}.
Eccentricites of $e\simeq 0.03$ can also be sustained by turbulence-driven stochastic
torques \citep{nel05}.

For massive gap-opening planets, the eccentricity damping timescale
given in Equation (\ref{eq:tau_ecc_2h}) is not valid. The evolution of the eccentricity is sensitive to 
the shape of the gap \citep[e.g.][]{moo08,tsa14}. 
In order to estimate $\tau_{e}$, one needs to know the disc surface
density distribution. Using the outputs of hydrodynamical simulations for the reference 
values of $e=0.1$, $q=10^{-4}$, $q_{\rm disc}=10^{-4}$, $h=0.05$ and
$\alpha=10^{-3}$, \citet{che21} obtain 
$\tau_{e}\simeq 200$ orbits, which is a factor $\sim 5$ shorter than predicted by
Equation (\ref{eq:tau_ecc_2h}).
On the other hand, \citet{bai21} argues that the eccentricity of sub-Jovian
planets might not always be  damped due to a complex flow pattern near the planet
that, however, can only be recovered correctly in high-resolution 3D simulations.

Numerical simulations of the evolution of planets with masses above
$1M_{J}$ indicate that the eccentricity can even grow \citep{pap01,gol03,kle06,dan06,moo08,
dun13,duf15b,tey16,rag17,rag18,ros17,mul19,bar21,deb21,leg21}.
 \citet{duf15b} argue that the planet eccentricity is 
expected to grow up to $\sim 0.1$, because for larger values the planet crashes 
with the walls of the gap. Interestingly, \citet{leg21} find cases where the eccentricity 
can grow up to $\simeq 0.25$ in their 2D low-viscosity simulations.

Eccentricities $\simeq 0.03-0.2$ could  
be sustained through mean motion resonances 
after the planets have been captured in resonance by disc-driven migration
\citep[e.g.,][]{kle04,pap05,ter07,pie08,lee09,mar10,mat10,bat15,dec15,izi17,kan20}, 
by repeated resonance crossings \citep{chi02}.

One of the recently proposed mechanisms that can prevent the damping of a planet's eccentricity is the interference of the main spiral arms of the planet with the spiral waves formed in the gas disc \citep[][]{ChCh2022}. These spiral waves propagating in the disc can be the result of the secondary arms formed by the planet itself \citep[see][and references therein]{zhu22} and can also be produced by a vortex that is far from the planet \citep[][]{ChCh2022}. The latter mechanism could maintain planets of $\lesssim 0.6M_{J}$ in eccentric orbits.

In summary, although there is currently a debate about how a planet can maintain its eccentricity, there is 
a number of observational motivations to explore the consequences of having planets in eccentric orbits \citep[][]{li19,bai21,RZ2021,che21,zhu22}. It seems unlikely that the assumption that planets are on circular orbits
applies to all planets in general. Indeed, eccentricities of $\sim h$ are likely in early stages of planet formation.

\section{Summary}
\label{sec:conclusions}
We investigated the depth and radial density profile of gaps carved by planets with eccentricities $e/h \leq 4$. 
By means of a semi-analytical model based
on the impulse approximation, we estimated the one-sided torque $T_{1s}$
exerted on the disc by the planet. We found that the strength of the one-sided torque
decreases with increasing eccentricity.
Invoking the zero-dimensional approximation,
we derived a scaling relation between the gap depth and $q$, $h$, $\alpha$, and $e$
(Eq. \ref{eq:gap_depth_obvious}).  By design, the scaling agrees with
the scaling of \citet{duf15a} for $e=0$ \citep[see also][]{fun14,kan15,kan17}.
The model, however, has a free function $\beta$ which was calibrated against
2D hydrodynamical simulations. Therefore, our approach was empirical.

We performed 2D hydrodynamical simulations where the planetary orbit was fixed.
The aspect ratios and viscosities of the simulated discs were $h\in[0.025,0.07]$ 
and $\alpha \in [10^{-4},10^{-3}]$, respectively. Our experiments indicate that 
our scaling relation (Eqs \ref{eq:fitT1s} and \ref{eq:gap_depth_obvious}) correctly recovers the gap depth up to moderately deep gaps ($\Sigma_{\rm gap}/\Sigma_{0,\rm gap}>10^{-2}$) and for eccentricities $\tilde{e}\leq 4$.
It was shown that the scaling relation provides fairly well the depth 
of the gap produced by planets that have reached their pebble isolation mass
in the low-viscosity regime ($\alpha\leq10^{-3}$). 

We calculated the radial profile of the gap in a model where the 
deposition of angular momentum of the waves is non-local by adopting
the wave dissipation rate derived by \cite{goo01} and \citet{raf02}.
When the torque is assumed to be injected in a small region around $R=a$,
the resulting gap depths become similar to simulations.
However, the gaps predicted 
are generally narrower than gaps in the simulations. On the other hand, if we
assume that the torque is excited along a radially extended region as predicted in
the impulse approximation, the widths of the gap are overestimated.
More sophisticated models are required to derive the deposition torque density
in order to determine the profile of gaps opened by eccentric objects.

\section*{Acknowledgements}
We are extremely grateful to the referee for thoughtful comments and insightful suggestions.
This work was supported by the Czech Science Foundation (grant 21-11058S).
The work of O.C. was supported by the Charles University Research program (No. UNCE/SCI/023). Computational resources were available thanks to the Ministry of Education, Youth and Sports of the Czech Republic through the e-INFRA CZ (ID:90140, LM2018140).

\section*{Data Availability}

The FARGO-ADSG code is available from \href{http://fargo.in2p3.fr/-FARGO-ADSG-}{http://fargo.in2p3.fr/-FARGO-ADSG-}. The input files for generating our hydrodynamical simulations will be shared on reasonable request to the corresponding author.

\appendix
\section{The excitation torque density}
\label{sec:explicit_form}
In this Appendix, we provide an explicit form for the excitation torque density
found in Section \ref{sec:tq_density}, namely 
\begin{equation}
\Lambda_{e}(x) = \Sigma(x) 
\int_{1-e}^{1+e}\mathcal{P} (\xi) \,\delta h_{e} \,|v_{\rm rel}|
\, d\xi.
\label{eq:torque_density0_app}
\end{equation}
We remind that  $x\equiv R-a$. 

\begin{figure*}
\includegraphics[width=150mm, height=140mm]{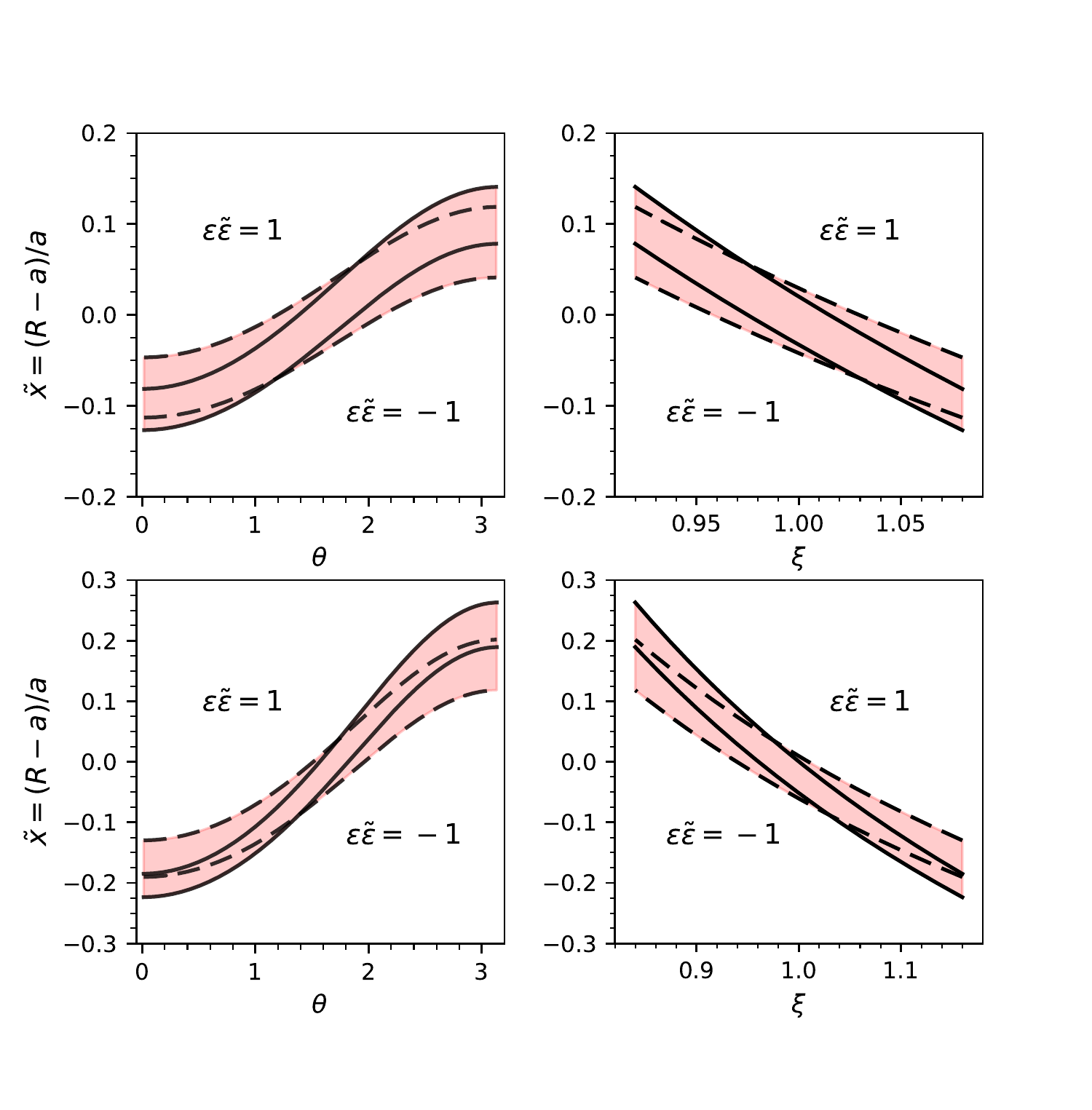}
\caption{Diagrams of $\varepsilon \tilde{\varepsilon}$ in the plane $(\theta,\tilde{x})$ (left
panels) and in the plane $(\xi,\tilde{x})$ (right panels) for $e=0.08$ (top panels) and
$e=0.16$ (bottom panels). In the red shaded region, $\varepsilon\tilde{\varepsilon}=0$. Above the
red region, $\varepsilon\tilde{\varepsilon}=1$. Below the red area, it is $-1$. In the region between 
the solid lines, $\varepsilon\tilde{\varepsilon}=0$ because the relative velocity is subsonic.
Between the dashed lines, $\varepsilon \tilde{\varepsilon}=0$ because the impact parameter is less
than $R_{\rm cut}$. We took $R_{\rm cut}=0.9H_{p}$.
}
\label{fig:eps_eps}
\end{figure*}

The change of angular momentum $\delta h_{e}$ given in Equation (\ref{eq:gas_scatter}) can be expressed as
\begin{equation}
\delta h_{e} = \frac{2\varepsilon \eta^{2}q^{2}\omega^{4} a^{7}}{\xi v_{\rm rel}^{3} \Delta^{2}}.
\label{eq:he}
\end{equation}
Here we have used that $G^{2}M_{p}^{2} = q^{2}\omega^{4}a^{6}$ and $R_{p}= \eta^{2} a\xi^{-1}$.
Substituting Equation (\ref{eq:he}) into Equation (\ref{eq:torque_density0_app}), we obtain
\begin{equation}
\Lambda_{e}(x) = 2\eta^{2} q^{2}\Sigma(x) \omega^{4}a^{7}
\int_{1-e}^{1+e}\xi^{-1}\mathcal{P} (\xi)
\frac{\tilde{\varepsilon}\varepsilon}{v_{\rm rel}^{2} \Delta^{2}} 
\, d\xi,
\label{eq:torque_density0}
\end{equation}
where $\tilde{\varepsilon}={\rm sign} \,v_{\rm rel}$.

To evaluate the integral in Eq. (\ref{eq:torque_density0}),  we write $\Delta$ 
and $v_{\rm rel}$ in terms of $x$ and $\xi$ as follows. From the
definition of $\Delta$, namely $\Delta=R-R_{p}$, it follows that
\begin{equation}
\Delta = a+x-R_{p} =  a\left(1-\frac{\eta^{2}}{\xi}\right)+x,
\label{eq:Delta_app}
\end{equation}
where we have used Equation (\ref{eq:Rp}).

Analogously, we can write $v_{\rm rel}$ as a function of $x$ and $\xi$ by
substituting Equation (\ref{eq:Delta_app}) into 
Equation (\ref{eq:vrel_mod}). The result is
\begin{equation}
v_{\rm rel} = \frac{1}{2} \eta^{-3}\omega a \xi^{3/2} \mathcal{A}(\xi),
\label{eq:vrel_app}
\end{equation}
with
\begin{equation}
\mathcal{A}(\xi)= 2\sqrt{\xi}-5 +3\eta^{-2}\left[1+\tilde{x}\right] \xi,
\end{equation}
where $\tilde{x}\equiv x/a$.

Substituting Eqs (\ref{eq:Delta_app}), (\ref{eq:vrel_app}) and (\ref{eq:time_fraction}) into 
Equation (\ref{eq:torque_density0}),
we finally obtain
\begin{equation}
\frac{\Lambda_{e}(\tilde{x})}{q^{2}\Sigma(\tilde{x})\omega^{2}a^{3}} = \frac{8\eta^{8}}{\mathcal{I}_{e}}\!\!
\int_{1-e}^{1+e}\!\!\frac{\xi^{-6}\tilde{\varepsilon}\varepsilon\,d\xi}{\sqrt{e^{2}-(\xi-1)^{2}} ({1-\eta^{2}\xi^{-1}+\tilde{x})^{2}}
\mathcal{A}^{2}}.
\label{eq:explicit_density_tq}
\end{equation}
This integral was performed numerically to obtain the excitation torque density as a function
of $\tilde{x}$, for a certain value of $e$.

In the circular case ($e=0$), the problem is axisymmetric and therefore $\varepsilon
\tilde{\varepsilon}$ does not depend either on $\theta$ or on $\xi$ 
(see Eq. \ref{eq:circular_case_tq} below).
In the general case, however, $\varepsilon\tilde{\varepsilon}$ depends 
on $\xi$ and $\tilde{x}$.
For illustration, Figure \ref{fig:eps_eps} shows the values 
of $\varepsilon\tilde{\varepsilon}$
in the $(\theta,\tilde{x})$ and $(\xi,\tilde{x})$ planes, for two values of $e$.

The torque density given in Equation (\ref{eq:circular_density_tq}) for the circular case
should be recovered in the limit $e\rightarrow 0$. In fact,
if $e\rightarrow 0$, then $\eta\rightarrow 1$, $\xi\rightarrow 1$, $\mathcal{I}_{e}=\pi$,
$\mathcal{A}\rightarrow 3\tilde{x}$ and therefore Equation 
(\ref{eq:explicit_density_tq}) simplifies to
\begin{equation}
\Lambda_{e=0}(x) =\frac{8q^{2}\Sigma \omega^{2}a^{3}}{\pi} \int_{0}^{\pi}
\frac{\varepsilon\tilde{\varepsilon}}{9\tilde{x}^{4}} \,d\theta
=\frac{8}{9} \varepsilon\tilde{\varepsilon}q^{2}\Sigma \omega^{2}a^{3} \left(\frac{a}{x}\right)^{4},
\end{equation}
with
\begin{equation}
\varepsilon\tilde{\varepsilon}= 
\left\{
\begin{array}{cl}
1 &\textup{if  }  \,\,x> {\rm max}\{R_{\rm cut}, 2H_{a}/3\}\\
  &  \\
-1 &\textup{if  }  \,\,x<-{\rm max}\{R_{\rm cut}, 2H_{a}/3\}\\
 & \\
0 & \textup{otherwise.}
\end{array}
\right.
\label{eq:circular_case_tq}
\end{equation}

\begin{figure*}
\includegraphics[width=200mm, height=60mm]{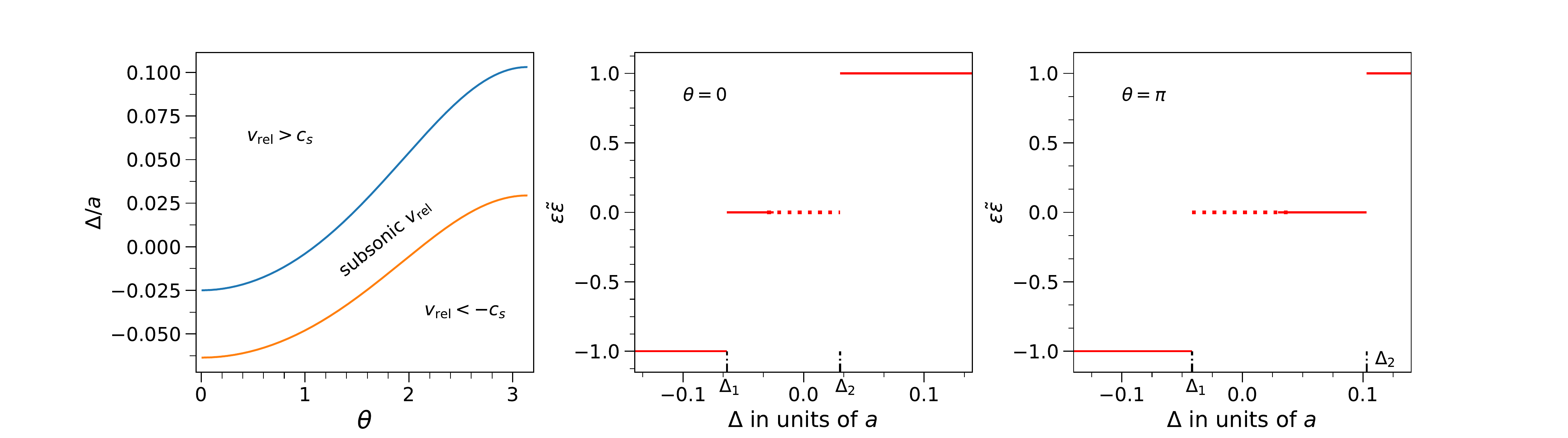}
\caption{Diagrams of the streamlines that contribute to the torque for a case with
$e=0.16$, $h=0.04$ and $R_{\rm cut}=0.9H_{p}$. The left figure delineates
the impact parameters for those streamlines that are supersonic. For impact 
parameters that lie in the region between the two curves, $v_{\rm rel}$ is subsonic 
($|v_{\rm rel}|<c_{s}$). The middle
and right panels show $\varepsilon\tilde{\varepsilon}$ as a function of $\Delta$ when
the perturber is at the pericentre ($\theta=0$) and at apocentre ($\theta=\pi$),
respectively. Streamlines with impact parameters having 
$\varepsilon\tilde{\varepsilon}=0$ do not contribute to the torque.
Dotted lines indicate those impact parameters for which $\varepsilon=0$
because $\Delta<R_{\rm cut}$.
}
\label{fig:varepsilon}
\end{figure*}

\section{One-sided torque calculation}
\label{sec:one_sided_tq_app}
In this Appendix, we describe the computation of 
the transfer rate of angular momentum to the external disc ($R>a$) or 
outer one-sided torque, denoted by $T_{1s}^{(o)}$. The derivation
of the inner one-sided torque is analogous.

$T_{1s}^{(o)}$ can obtained by integrating $\Lambda_{e} (x)$, given in Equation
(\ref{eq:torque_density0}) over the impact parameter:
\begin{equation}
T_{1s}^{(o)}= 2\eta^{2} q^{2}\Sigma_{a} \omega^{4}a^{7}
\int_{1-e}^{1+e}\xi^{-1}\mathcal{P} (\xi)\int_{a-R_{p}}^{\infty} 
\frac{\tilde{\varepsilon}\varepsilon}{v_{\rm rel}^{2} \Delta^{2}} 
\,d\Delta\, d\xi,
\label{eq:torque_ext}
\end{equation}
where we have approximated $\Sigma(x)\simeq \Sigma_{a}$, the surface density at $R=a$.

Since $\tilde{\varepsilon} \varepsilon$ is a piecewise constant function
that takes values $-1,0$ or $1$, we can
compute the integral over $\Delta$ analytically by splitting the integral up into sum 
of integrals on subintervals, being 
$\varepsilon=\tilde{\varepsilon}=1$, $\varepsilon=-\tilde{\varepsilon}=1$ or
$\tilde{\varepsilon}\varepsilon=0$, along
each subinterval:
\begin{equation}
\int_{a-R_{p}}^{\infty} 
\frac{\tilde{\varepsilon}\varepsilon}{v_{\rm rel}^{2} \Delta^{2}}\, d\Delta=
\tilde{\varepsilon}_{i}
\int_{\Delta_{i}}^{\Delta_{i+1}} \frac{d\Delta}{v_{\rm rel}^{2} \Delta^{2}}+
\tilde{\varepsilon}_{i+1}
\int_{\Delta_{i+1}}^{\Delta_{i+2}} \frac{d\Delta}{v_{\rm rel}^{2} \Delta^{2}}+...
\label{eq:integral_split}
\end{equation}
where $\tilde{\varepsilon}_{j}$ is the value of $\tilde{\varepsilon}$ in the
interval $[\Delta_{j},\Delta_{j+1}]$.
The impact parameters $\Delta_{j}$ satisfy that either 
$|\Delta_{j}|=R_{\rm cut}$ or
the relative velocity for streamlines with those impact parameters is $c_{s}$.
For the simplest circular case, $\tilde{\varepsilon}\varepsilon$ is given in 
Equation (\ref{eq:circular_case_tq}) (just note that $\Delta=x$ in this case).
Examples of the different subintervals for $e=0.16$ are shown in Figure \ref{fig:varepsilon}. 
In the cases shown, we have two subintervals where 
$\varepsilon\tilde{\varepsilon}\neq 0$,
but for larger eccentricities we may have three subintervals.

Substituting the value of $v_{\rm rel}$ from Equation (\ref{eq:vrel_mod}) 
and integrating over $\Delta$, it is simple to show that, for $e\neq 0$, 
each integral in the right-hand-side of Equation (\ref{eq:integral_split}) can be written as
\begin{equation}
\int_{\Delta_{j}}^{\Delta_{j+1}} \frac{d\Delta}{v_{\rm rel}^{2} \Delta^{2}}
 = \frac{3\eta^{4}}{2 \omega^{2} a^{3}}
\frac{1}{\xi^{3/2}\hat{\xi}^{3}} \left[f_{\xi}(\alpha_{j+1})-f_{\xi}(\alpha_{j})\right],
\label{eq:integral_Delta}
\end{equation}
where $\hat{\xi}\equiv \xi-\sqrt{\xi}$,
\begin{equation}
\alpha_{j} = \frac{3\xi^{3/2}}{2\eta^{2}\hat{\xi}} \frac{\Delta_{j}}{a},
\label{eq:alphaj}
\end{equation}
and
\begin{equation}
f_{\xi}(\alpha) = 2\ln\left(\frac{1+\alpha}{\alpha}\right) -\frac{1}{\alpha}-\frac{1}{1+\alpha}.
\end{equation}
If $e=0$, the respective integrals are
\begin{equation}
\int_{\Delta_{j}}^{\Delta_{j+1}} \frac{d\Delta}{v_{\rm rel}^{2} \Delta^{2}}
 = \frac{1}{2\omega^{2}a^{3}}\left(\frac{1}{\alpha_{j}^{3}}-\frac{1}{\alpha_{j+1}^{3}}\right).
\end{equation}

In order to compute $T_{1s}^{(o)}$ we carry out the following procedure.
For a given $\xi$, we find out the values of $\Delta_{j}$.  Once we know $\Delta_{j}$,
we derive the respective $\alpha_{j}$ from Equation (\ref{eq:alphaj}). Then, we
evaluate the integrals in Equation (\ref{eq:integral_split}) using Equation (\ref{eq:integral_Delta}). We repeat the procedure for an equally spaced set of
values of $\xi$, and numerically integrate Equation (\ref{eq:torque_ext}) over $\xi$.

\section{The condition for positive solutions in a simple analytical case}
\label{sec:analytical_diff_eq}
In order to illustrate under which conditions, Equation (\ref{eq:nonlocal_deposition}) has solutions
with negative values of $\tilde{\Sigma}$ (non-physical solutions), we consider a
simple analytical model where the density torque is given by piecewise constant
function as follows
\begin{equation}
\frac{\Lambda_{e}(b)}{\Sigma_{0,a}\omega^{2}a^{3}}
=
\left\{
\begin{array}{cl}
q^{2}\tilde{\Sigma} \Lambda_{c} &\textup{if  }  \,\,1<b < b_{\rm max} \\
  &  \\
-q^{2}\tilde{\Sigma}\Lambda_{c} &\textup{if  }  \,\, -b_{\rm max}<b<1\\
 & \\
0 & \textup{otherwise,}
\end{array}
\right.
\label{eq:analytical_model}
\end{equation}
where $\Lambda_{c}$ is a positive constant and $b_{\rm max}$ is a certain maximum distance. For simplicity, we further assume that $\Lambda_{\rm dep}=\Lambda_{e}$.
Then, Equation (\ref{eq:nonlocal_deposition}) in the outer disc (at $1<b<b_{\rm max}$)
has the form
\begin{equation}
\frac{d}{db}(b f(b) \tilde{\Sigma}) -\left(\frac{2q^{2}\Lambda_{c}}{3\pi \nu_{0}}\right) b\tilde{\Sigma}=F_{0}.
\label{eq:local_deposition_analytical_model}
\end{equation}
We now specialize to $F_{0}=1$ and constant $\alpha$ parameter (i.e. $\nu_{0}=\alpha h^{2}$ and $f(b)=b$), the equation for $\tilde{\Sigma}$ becomes
\begin{equation}
\frac{d}{db}(b^{2} \tilde{\Sigma}) -2\mathcal{A} b\tilde{\Sigma}=1.
\label{eq:local_deposition_analytical_model_2}
\end{equation}
where $\mathcal{A}$ is the only parameter of the problem and is given by
\begin{equation}
\mathcal{A}=\frac{q^{2}\Lambda_{c}}{3\pi \alpha h^{2}}.
\end{equation}
For $\mathcal{A}\neq 1/2$, the solution of Equation (\ref{eq:local_deposition_analytical_model_2}) with the condition that
$\tilde{\Sigma}(b_{\rm max})= 1/b_{\rm max}$ is
\begin{equation}
\tilde{\Sigma} (b)= \frac{1}{(1-2\mathcal{A})b}\left(1-2\mathcal{A} 
\left(\frac{b}{b_{\rm max}}\right)^{2\mathcal{A}-1}\right).
\end{equation}
Therefore $\tilde{\Sigma}=0$ at a critical radius $b_{c}$ given by
\begin{equation}
b_{c}=(2\mathcal{A})^{1/(1-2\mathcal{A})}b_{\rm max}.
\end{equation}
The condition for the solution to be positive is $b_{c}<1$, which implies
\begin{equation}
(2\mathcal{A})^{1/(1-2\mathcal{A})} < \frac{1}{b_{\rm max}}.
\end{equation}
For instance, for $b_{\rm max}=1.1$, this condition implies $\mathcal{A}<19.5$.

\section{Deposition torque density in a model with a spatially-extended
excitation torque} 
\label{sec:appD}
In order to determine $\Lambda_{\rm dep}$,
we will assume that the positive torque excited at $R>R_{\Lambda}$ leads to the formation of waves 
that carry (positive) angular momentum outwards, whereas the negative torque injected at $R<R_{\Lambda}$ excites
waves that carry (negative) angular momentum
inwards. Focusing on the region $R>R_{\Lambda}$ first,
the deposition torque at a radius $R$ is determined by the sum of the 
contribution of all the waves
that damp at that radius, i.e. 
\begin{equation}
\Lambda_{\rm dep}(R) = \int_{R_{\Lambda}}^{R} \Lambda_{\rm e} (r_{\rm exc}) \bigg|\frac{d\Phi}{dR}\bigg|_{(r_{\rm exc},R)} dr_{\rm exc},
\label{eq:Lambda_dep}
\end{equation}
where $|d\Phi/dR|$ evaluated at $(r_{\rm exc}, R)$ is the fraction of the angular momentum 
that excited at $r_{\rm exc}$ is deposited at $R$.

Previous works have derived $\Phi$ for the density waves generated by planets in 
circular orbits in the weakly nonlinear regime \citep{goo01,raf02,gin18}.
For eccentricities $e\leq 2h$, \citet{zhu22}
argue that the structure of spiral arms is rather similar to the circular case and it is likely that
the deposition of angular momentum is also similar.
Although it remains unclear how waves launched by planets with $e>2h$ are damped, we adopt
a simple generalization of the angular momentum deposition found by \citet{raf02} in the case of
circular orbits.

More specifically, focusing again on $R>R_{\Lambda}$, we will assume that 
the dimensionless 
angular momentum flux, $\Phi (r_{\rm exc}; R)$, is
\begin{equation}
\Phi(r_{\rm exc}; R)=
\left\{
\begin{array}{cl}
1 &\textup{if  }  \,\,\tau(r_{\rm exc}; R)<\tau_{\rm sh} \,\,\,\textup{and } R>r_{\rm exc}\\
  &  \\
\sqrt{\tau_{\rm sh}/\tau} &\textup{if  }  \,\,  \tau(r_{\rm exc}; R)>\tau_{\rm sh} \,\,\,\textup{and } R>r_{\rm exc}\\
&\\
0 & \textup{otherwise,}
\end{array}
\right.
\label{eq:damping_Rafikov}
\end{equation}
where
\begin{equation}
\tau(r_{\rm exc}; R) = \frac{3}{(2 h^{2})^{5/4}} \int_{1}^{R/r_{\rm exc}} (s^{3/2}-1)^{3/2}
s^{(5p_{2}+p_{1})/2-11/4} ds 
\label{eq:rafikov43}
\end{equation}
where $p_{1}$ and $p_{2}$ represent the same power indexes as in Section \ref{sec:shape_gap}.

Note that Equation (\ref{eq:rafikov43}) 
corresponds to Equation (43) in \citet{raf02}. The only difference is that waves are not only excited at $r_{\rm exc}\simeq a$. Thus, Equations
(\ref{eq:damping_Rafikov}) and (\ref{eq:rafikov43}) show explicitly the dependence on $r_{\rm exc}$.
We will omit the expressions of $\Lambda_{\rm dep}$ for $R<R_{\Lambda}$ because the derivation is similar.

\bsp	\label{lastpage}
\end{document}